\documentclass[3p,10pt]{elsarticle}

\usepackage{amssymb}
\usepackage{amsmath}
\usepackage{amsthm}
\usepackage{amsfonts}
\usepackage{graphicx,epstopdf}
\usepackage{dcolumn}
\usepackage{bm}
\usepackage{mathtools}

\DeclareMathOperator{\Span}{span}

\DeclareMathOperator{\per}{per}

\DeclarePairedDelimiter\floor{\lfloor}{\rfloor}

\journal{Physica A}

\begin{document}

\begin{frontmatter}

\title{Hartree-Fock approximation for bosons with symmetry-adapted variational wave functions}
\author[1,2]{B. R. Que}
\ead{BiaoRui.Que@gmail.com}
\author[1,2]{J. M. Zhang}
\ead{wdlang06@163.com}
\author[3]{H. F. Song}
\ead{song\_haifeng@iapcm.ac.cn}
\author[3]{Y. Liu}
\ead{liu\_yu@iapcm.ac.cn}
 \affiliation[1]{organization={Fujian Provincial Key Laboratory of Quantum Manipulation and New Energy Materials,
College of Physics and Energy, Fujian Normal University},
            city={Fuzhou},
            country={China}}
\affiliation[2]{organization={Fujian Provincial Collaborative Innovation Center for Optoelectronic Semiconductors and Efficient Devices},
          city={Xiamen},
            country={China}}
\affiliation[3]{organization={Laboratory of Computational Physics, Institute of Applied Physics and Computational Mathematics},
            city={Beijing},
            country={China}}

\begin{abstract}
The Hartree-Fock approximation for bosons employs variational wave functions that are a combination of permanents. These are bosonic counterpart of the fermionic Slater determinants, but with the significant distinction that the single-particle orbitals used to construct a permanent can be arbitrary and do not need to be orthogonal to each other. Typically, the variational wave function may break the symmetry of the Hamiltonian, resulting in qualitative and quantitative errors in physical observables.  A straightforward method to restore symmetry is projection after variation, where we project the variational wave function onto the desired symmetry sector. However, a more effective strategy is variation after projection, which involves first creating a symmetry-adapted variational wave function and then optimizing its parameters.  We have devised a scheme to realize this strategy and have tested it on various models with symmetry groups ranging from $\mathbb{Z}_2$, $\text{C}_L$, to $\text{D}_L$.  In all the models and symmetry sectors studied, the variational wave function accurately estimates not only the energy of the lowest eigenstate but also the single-particle correlation function, as it approximate the target eigenstate very well on the wave function level. We have applied this method to study few-body bound states, superfluid fraction, and Yrast lines of some Bose-Hubbard models. This approach should be valuable for studying few-body or mesoscopic bosonic systems. 
\end{abstract}



\end{frontmatter}


\section{Introduction}

The Hartree-Fock approximation is widely used in atomic, nuclear, and condensed matter physics \cite{hartree, slater1,  gaunt,fock,  slater2, ostlund}. Although often labelled as a mean-field or self-consistent field theory, it is best understood as a variational method. The trial wave function involved is a Slater determinant, which is arguably the simplest wave function of an $N$-fermion system.  Its construction is as follows: Take $N$ single-particle orbitals $\{ \phi_i , 1\leq i \leq N \}$, form the product state $\phi_1(x_1)\phi_2(x_2) \ldots \phi_N (x_N )$, and then anti-symmetrizes it to fulfil the anti-symmetry condition. The resultant wave function can be written compactly as a determinant. Explicitly, with an appropriate normalization factor, the wave function is
\begin{eqnarray}\label{slater}
  \Xi(x_1,x_2, \ldots, x_N)   = \frac{1}{\sqrt{N!}} \sum_{P \in S_N } (-1)^P \phi_{P_1}(x_1) \phi_{P_2}(x_2) \ldots  \phi_{P_N}(x_N) = \frac{1}{\sqrt{N!}}\det (\phi_{i} (x_{j} )),
\end{eqnarray}
where $S_N$ is the permutation group. 

From the variational point of view, it should be straightforward to develop a Hartree-Fock approximation for bosons.
To construct a variational wave function, we again start from $N $ single-particle orbitals $\{ \phi_i , 1\leq i \leq N \}$. The only but essential difference with the fermionic case is that, after forming the product state  $\phi_1(x_1)\phi_2(x_2) \ldots \phi_N (x_N )$, we symmetrize it instead of anti-symmetrizing. The resulting variational wave function is then 
\begin{eqnarray}\label{permstate}
	\Phi(x_1,x_2, \ldots, x_N)   = \frac{1}{\sqrt{N!}} \sum_{P \in S_N }  \phi_{P_1}(x_1) \phi_{P_2}(x_2) \ldots  \phi_{P_N}(x_N) = \frac{1}{\sqrt{N!}} \per (\phi_{i} (x_{j} )),
\end{eqnarray}
and is referred to as a permanent state. Once we have the variational wave function, the rest is to optimize the orbitals to minimize the energy expectation value. Here in passing, we should mention that the well-known Gross-Pitaevskii approximation \cite{gross,pitaevskii} also employs a permanent variational wave function, albeit a very special one, in which all the orbitals are clamped to be identical. While this condensate-type wave function works well for weakly interacting bosons \cite{rmp1,rmp2}, it becomes inadequate for  intermediate or strong interactions, such as in Tonks-Girardeau gases \cite{girardeau,weiss}. Therefore, it is crucial to allow the orbitals to vary independently and freely.   

Simple as it is, the scheme above is rarely carried out in practice.  As far as we know, so far there are only five papers \cite{ igor1, igor2, martin1,martin2, zhang4} that earnestly use a permanent wave function as a variational wave function for a bosonic system. The reason is that permanents are much more difficult to handle than determinants.

First of all, while determining the norm of an arbitrary determinant state is quite easy, determining that of an arbitrary permanent state could be quite challenging. It involves calculating the determinant and permanent of the Gram matrix $A_{ij} = \langle \phi_i | \phi_j \rangle$ of the single-particle orbitals, respectively. That is, 
\begin{eqnarray}
		\langle \Xi |\Xi \rangle = \det(A) , \quad 	\langle \Phi |\Phi \rangle = \per(A ) . 
\end{eqnarray}
While the determinant can be computed in polynomial time by Gaussian elimination, it is generally believed that the permanent cannot be computed in polynomial time. Actually, for fermions, the single-particle orbitals can always be orthonormalized beforehand (say, via the Gram-Schmidt procedure), and then the Slater determinant is automatically normalized to unity. However, for bosons, there is no constraint on the single-particle orbitals, and calculating the permanent of a generic positive semi-definite matrix $A$ is inevitable. This causes great difficulties. 

Theoretically, the Hartree-Fock theory for bosons is far less elegant than that for fermions. For fermions, the optimal single-particle orbitals are eigenstates of the so-called Fock operator, which is a single-particle Hamiltonian consisting of an effective single-particle potential. This effective single-particle potential, although generally non-local, is the ``mean-field'' in the theory. However, for bosons, we no longer have a Fock operator or any counterpart of it. Each of the optimal single-particle orbitals is a solution of a generalized eigenvalue problem, which varies from orbital to orbital \cite{comment1}. The mean-field picture is lost. 
 
Practically, that we have to calculate the permanent and the best algorithm \cite{ryser,fastryser} for this purpose has a time-complexity of $O( 2^{N-1}  N)$ means that we cannot handle very large $N$. So far, the largest value of $N $ that has been treated is 12. Therefore, the variational approach based on permanent wave functions is applicable not to genuine many-body systems, but only to few-body or mesoscopic systems \cite{review3, review2, review1}. 

Despite of these difficulties, our previous work \cite{zhang4} shows that permanents can be highly expressive. For instance, for the one-dimensional Bose-Hubbard model with periodic boundary condition and at unit filling, we find that when the system is as large as with $N=12$ particles, the overlap between the exact ground state and the optimal permanent is at least 0.96 throughout the range of the model parameters. This is particularly striking given that the dimension of the many-body Hilbert space is as large as $1\,352\,078$, while the permanent variational wave function contains only 12 parameters (representing a single orbital). Extrapolating these results to $N=20$, we infer that the overlap between the exact ground state and the optimal permanent remains at least 0.9, even though the dimension of the many-body Hilbert space now grows exponentially to approximately $6.89\times 10^{10}$. Meanwhile, the permanent wave function still only involves 20 parameters. 

Here we are talking about the overlap rather than the more commonly considered ground state energy, as overlap is a more stringent criterion than energy for accessing the accuracy of a variational wave function. In a multi-body system with many degrees of freedom, it is possible for the trial wave function and the exact ground state to have very similar energies, while their overlap is vanishingly small. Anyway, for a multi-body system, often the first excited state is very close to the ground state in energy, yet its overlap with the latter is zero. The key point is that a multi-body wave function has more or less a tensor structure, and the overlap depends, roughly speaking, multiplicatively  on each degree of freedom, while the energy depends additively on each degree of freedom. Therefore, for two wave functions to have a significant overlap, they must be similar in each degree of freedom. In contrast, the condition for closeness in energy is much less strict---a modification in just one degree of freedom can cause the overlap to diminish substantially, while the energy is only affected $O(1/N )$. In summary, closeness in overlap generally implies closeness in energy, but the reverse is not necessarily true. In the Bose-Hubbard model discussed earlier, due to the high overlap between the permanent state and the exact ground state, the permanent state accurately reproduces not only the ground state energy but also other quantities such as the single-particle correlation function. 
 
We would argue that the high overlap between the permanent state and the exact ground state is, to some extent, a fortuitous consequence of the Bose symmetry. A bosonic wave function, unlike a fermionic one, can be positive everywhere. Indeed, for the Bose-Hubbard model, by the Frobenius-Perron theorem \cite{horn, bhatia}, it is easily proven that the ground state is positive throughout the real space \cite{ed}. Now a permanent state made of single-particle orbitals that are positive everywhere is positive everywhere too. Consequently, in calculating the overlap between the exact state and the permanent state, only positive values accumulate, which is of course favorable for achieving a large value ultimately. In contrast, fermionic wave functions necessarily alternate in sign, which likely leads to cancellations between positive and negative contributions in overlap calculations. Therefore, in this regard, the Hartree-Fock approximation is expected to perform better for bosonic systems than for fermionic ones.
 
We thus think it is worthwhile to develop the Hartree-Fock theory for bosons further. In this paper, we tackle the symmetry breaking problem of the Hartree-Fock approximation. It is quite common that the Hartree-Fock solution of the ground state breaks the symmetry of the Hamiltonian. In atomic physics, we have the spin contamination  problem \cite{ostlund}, where the Hartree-Fock wave function is not an eigenstate of the total spin-squared operator. In condensed matter physics, for the jellium model, we have the famous Overhauser instability \cite{overhauser}, where the most symmetric solution, i.e., the Fermi sea solution is unstable with respect to spin or charge modulation. In nuclear physics, the discovery that some nuclei are deformed is a Nobel prize-winning finding \cite{bohr,butler,frauendorf}. Sometimes, the symmetry breaking of the Hartree-Fock solution is a feature---it is physically meaningful and reflects the intrinsic structure of the system. However, in many cases, it is just a nuisance. Therefore, to get physical quantities qualitatively correct and quantitatively accurate, we have to restore the symmetry of the wave function.

In our previous work \cite{zhang4}, we were only concerned with the global ground state, and the strategy we took was projection after variation (PAV) \cite{ring}. We simply project the variational state onto the desired symmetry sector. For the ground state, which often belongs to the trivial representation, we simply take the average of the states generated by the symmetry operators. However, a better approach is variation after projection (VAP) \cite{ring, symmetry1, symmetry2}. In this approach, a trial wave function satisfying the symmetry requirement is constructed first and then its parameters are optimized.  Although this scheme might be difficult to implement in other problems, in our study of lattice models, it is perfectly compatible with the existing framework and can be readily implemented. The advantage of the variation after projection approach to the projection after variation approach is that, it is not only  more accurate (by construction), but more importantly, it is more suitable for targeting the lowest eigenstate in an arbitrary symmetry sector. 

This paper is organized as follows. In Sec.~\ref{secalg}, we introduce the numerical algorithm. We shall first review the greedy algorithm introduced in \cite{zhang4} to construct an optimal multi-configuration wave function, and then discuss how to implement the variation after projection scheme. Section \ref{secbhm} contains many case studies. We shall study the Bose-Hubbard model on various lattices and with various boundary conditions, so that various symmetry groups can be realized. Specifically, we study one-dimensional lattices with  the open boundary condition ($\mathbb{Z}_2$ symmetry), one-dimensional lattices with the periodic boundary condition ($\text{C}_L$ or $\text{D}_L$ symmetry), and two-dimensional square lattices with the open boundary condition ($\text{D}_4$ symmetry). We shall also study the Bose polaron problem on a one-dimensional lattice with the periodic boundary condition. In all cases studied, we find that the symmetry-adapted permanent variational wave function can achieve high overlap with the exact state, accurately deliver the energy and the single-particle density matrix. We conclude with a discussion of the shortcomings of the current algorithm and future work in Sec.~\ref{secconclude}. 

\section{Numerical algorithm}\label{secalg}

Let us first introduce some notations. 

In this paper, we consider only lattice models. We have $N $ spinless bosons on a lattice of $L $ sites. The single-particle Hilbert space is 
\begin{eqnarray*}
	\mathcal{H} = \Span\{|x\rangle, 1\leq x \leq L  \},
\end{eqnarray*}
where $|x\rangle $ denotes the single-particle basis state on site $x$. The associated creation (annihilaton) operator is denoted as $ \hat{a}_x^\dagger $ ($ \hat{a}_x $). An arbitrary single-particle state or single-particle orbital $|\phi \rangle $ can be expanded with respect to this basis, $|\phi \rangle = \sum_x |x \rangle \langle x | \phi \rangle = \sum_x \phi(x) |x \rangle $. The associated creation operator is $ \hat{a}_\phi^\dagger = \sum_x \phi(x) \hat{a}_x^\dagger $. We emphasize that in this paper  an orbital is generally not normalized. 

The $N$-body Hilbert space is spanned by the orthonormal Fock states 
\begin{eqnarray}
	|\textbf{n} \rangle= \frac{(\hat{a}_1^\dagger)^{n_1}(\hat{a}_2^\dagger)^{n_2}\ldots (\hat{a}_L^\dagger)^{n_L}   }{\sqrt{ n_1! n_2 ! \ldots n_L ! }} |vac\rangle .
\end{eqnarray}
where $\textbf{n} \equiv (n_1, n_2 , \ldots, n_L )$ is an $L$-tuple with $n_x \geq 0 $ and $\sum_{x=1}^L n_x = N $. The number of such Fock states or the dimension of the $N$-boson Hilbert space is
\begin{eqnarray}\label{dimh}
	\mathcal{D} = \binom{N+L -1}{N} = \frac{(N+L-1)!}{N!(L-1)!}.
\end{eqnarray}
The Fock states are the basis that we use in the exact diagonalization (ED) calculations below. 

A generic Hamiltonian for the bosons on the lattice is of the form 
\begin{eqnarray}\label{1st}
	\hat{H} =  \sum_{i=1}^N \hat{K}(i ) + \sum_{1\leq i < j \leq N} \hat{W}(i,j ) .
\end{eqnarray}
Here $\hat{K}(i)$ is the kinetic and potential energy of the $i$th particle, while $\hat{W}(i, j )$ is the interaction energy between the $i$th and $j$th particle. Here we write the Hamiltonian in the first-quantization formalism because this is most convenient for our analytic derivations. However, in the following, we shall generally write the Hamiltonian in the second-quantization formalism for compactness. It should be easy to convert it back to the first-quantization formalism. For example, the Hamiltonian for the Bose-Hubbard model on a one-dimensional lattice with the periodic boundary condition is usually written as (note that here and henceforth we set the hopping strength to unity, so that the on-site interaction $U$ actually refers to a ratio and thus is a dimensionless quantity)
 \begin{eqnarray}
	\hat{H} &=& - \sum_{x=1}^{L} (\hat{a}_x^\dagger \hat{a}_{x+1} + \hat{a}_{x+1}^\dagger \hat{a}_{x})   + \frac{U }{2} \sum_{x=1}^{L}  \hat{a}_x^\dagger \hat{a}_x^\dagger \hat{a}_x \hat{a}_x .\quad
\end{eqnarray}
We can  read off the matrix-elements of the operators $\hat{K}$ and $ \hat{W} $ as 
\begin{subequations}\label{kandu}
\begin{eqnarray}
	\hat{K}_{x,x'}  &=& -(\delta_{x, x'+1} + \delta_{x, x'-1}) ,  \\
	\hat{W}_{x_1 x_2, x_1' x_2'} &=& U \delta_{x_1 , x_2} \delta_{x_1 , x_1'} \delta_{x_1 , x_2'} .  \label{kandu}
\end{eqnarray}
\end{subequations}

\subsection{Greedy algorithm}

Let us first review the greedy algorithm in Ref.~\cite{zhang4}, which was also partially developed in Ref.~\cite{martin2}. 

We shall use the notation $\hat{\mathcal{S}}(\phi_1, \phi_2, \ldots, \phi_N) $ to denote a permanent state constructed out of the orbitals $\{ \phi_i(x) , 1\leq i \leq N \}$ according to the product and symmetrization procedure in (\ref{permstate}). As proven in Ref.~\cite{zhang4}, a permanent state will be nonzero, i.e., $\langle \Phi |\Phi\rangle \neq 0 $, as long as the single-particle orbitals are all nonzero. 


To achieve high accuracy, we generally have to use a multi-configurational variational wave function, that is, a wave function as the sum of several permanents. The variational wave function is then of the form, 
\begin{eqnarray}\label{mconf}
	\Phi  &=& \sum_{\alpha = 1 }^M \Phi^{(\alpha )} =  \sum_{\alpha = 1 }^M \hat{\mathcal{S}}(\phi^{(\alpha)}_1, \ldots,\phi^{(\alpha)}_N ).
\end{eqnarray}
Here we have $M $ sets of single-particle orbitals. The object is to minimize the expectation value of the energy 
\begin{eqnarray}\label{ratio}
	E  = \frac{\langle \Phi |\hat{H}  | \Phi \rangle }{\langle \Phi | \Phi \rangle } = \frac{\sum_{\alpha,\beta= 1}^M \langle \Phi^{(\alpha) } | \hat{H} | \Phi^{(\beta)} \rangle }{\sum_{\alpha,\beta= 1}^M \langle \Phi^{(\alpha)} | \Phi^{(\beta)} \rangle }.
\end{eqnarray}
To tackle this multi-variable optimization problem, we take the greedy strategy. Let us fix the orbitals $\phi_{2\leq i \leq N }^{(1\leq \alpha \leq M)}$ and try to find optimal $\phi_1^{(1\leq \alpha \leq M )}$. The key observation is that, with $\phi_{2\leq i \leq N }^{(1\leq \alpha \leq M )}$ fixed, the $(\alpha \beta )$-th term in the numerator, $\langle \Phi^{(\alpha) } | \hat{H}  | \Phi^{(\beta)} \rangle$, and that in the denominator, $\langle \Phi^{(\alpha)} | \Phi^{(\beta)} \rangle $, are both  quadratic forms of $\phi_1^{(\alpha ) }$ and $\phi_1^{(\beta)}$.  That is, there exist \emph{single-particle} operators $\hat{F}^{(\alpha \beta )}$ and $\hat{G}^{( \alpha \beta )}$, which depend on the orbitals $\phi_{2\leq i \leq N }^{(\alpha )}$ and  $\phi_{2\leq i \leq N }^{(\beta )}$,  such that $\langle \phi_1^{(\alpha)} | \hat{F}^{(\alpha \beta )}  | \phi_1^{(\beta) } \rangle = \langle \Phi^{(\alpha)} |\hat{H} | \Phi^{(\beta)} \rangle$, $\langle \phi_1^{(\alpha)} | \hat{G}^{(\alpha \beta )}  | \phi_1^{(\beta )} \rangle = \langle \Phi^{(\alpha )} | \Phi^{(\beta)} \rangle$.
Because $ \langle \Phi^{(\alpha )} | \Phi^{(\beta)} \rangle = \langle \Phi^{(\beta )} | \Phi^{(\alpha)}\rangle^* $, we have $\langle \phi_1^{(\alpha)} | \hat{G}^{(\alpha \beta )}  | \phi_1^{(\beta )} \rangle = \langle \phi_1^{(\beta)} | \hat{G}^{(\beta \alpha  )}  | \phi_1^{(\alpha )} \rangle^*$ for arbitrary orbitals $\phi_1^{(\alpha )}$ and $\phi_1^{(\beta )}$, which means $\hat{G}^{(\beta \alpha  )} $ is the hermitian conjugate of $\hat{G}^{(\alpha \beta  )}$, i.e., $(\hat{G}^{(\alpha \beta)})^\dagger = \hat{G}^{(\beta \alpha) } $. Similarly, we have $	(\hat{F}^{(\alpha \beta)})^\dagger = \hat{F}^{(\beta \alpha)} $. 

We can then rewrite the ratio in (\ref{ratio}) as 
\begin{eqnarray}\label{ratio2}
	E  = \frac{\sum_{\alpha,\beta= 1}^M\langle \phi_1^{(\alpha)} | \hat{F}^{(\alpha \beta )}  | \phi_1^{(\beta) } \rangle }{\sum_{\alpha,\beta= 1}^M \langle \phi_1^{(\alpha)} | \hat{G}^{(\alpha \beta )}  | \phi_1^{(\beta) } \rangle }= \frac{\langle \phi_1 | \hat{F}  | \phi_1 \rangle }{\langle \phi_1 | \hat{G}  | \phi_1 \rangle }.
\end{eqnarray}
Here in the last equality,  we have concatenated the $M $ orbitals $\phi_1^{(\alpha)}$ into a vector of length $L M $, $\phi_1 =  (\phi_1^{(1)}; \phi_1^{(2)} ; \ldots ; \phi_1^{(M)} ) $, and have formed the $LM\times LM $ block matrices $\hat{F} \equiv  (\hat{F}^{(\alpha \beta )})_{1\leq \alpha, \beta \leq M} $ and $\hat{G} \equiv  (\hat{G}^{(\alpha \beta )})_{1\leq \alpha, \beta \leq M }$. As $\langle \phi_1 | \hat{F}  | \phi_1 \rangle = \langle \Phi |H| \Phi \rangle $ must be real, and $\langle \phi_1 | \hat{G}  | \phi_1 \rangle=  \langle \Phi | \Phi \rangle  $ must be nonnegative, we know $\hat{F}$ and $\hat{G}$ must be hermitian, and $\hat{G}$ even positive semi-definite. In practice, in our numerical calculations, we find that as long as $N\geq 3 $, $\hat{G}$ is always positive definite \cite{n2case}. 

Explicit expressions of the blocks $\hat{F}^{(\alpha \beta )}$ and $\hat{G}^{(\alpha \beta )}$ are given in the Appendix of Ref.~\cite{zhang4}. Here for illustration, we reproduce the derivation of $\hat{G}^{(\alpha \beta )}$, which is much simpler than $\hat{F}^{(\alpha \beta )}$. By definition, $\langle \phi_1^{(\alpha)} | \hat{G}^{(\alpha \beta )}  | \phi_1^{(\beta )} \rangle = \langle \Phi^{(\alpha )} | \Phi^{(\beta)} \rangle$. It is easily verified that this overlap is the permanent of the $N\times N $ Gram matrix $A_{ij}^{(\alpha \beta )} = \langle \phi_i^{(\alpha)} | \phi_j^{(\beta )} \rangle $, i.e., $\langle \Phi^{(\alpha )} | \Phi^{(\beta )} \rangle = \per(A^{(\alpha \beta )} )$. Now by an analogy of the Laplace expansion formula for the determinant, we have 
\begin{eqnarray}
	 \per(A^{(\alpha \beta )} ) &=&   \langle \phi_1^{(\alpha )} | \phi_1^{(\beta )} \rangle \per(A^{(\alpha \beta )};1|1 ) + \sum_{j_1 =2 }^N \langle \phi_1^{(\alpha )} | \phi_{j_1}^{(\beta )} \rangle \per(A^{(\alpha \beta )}  ;1|j_1 ) \nonumber \\
	&=& \langle \phi_1^{(\alpha )} | \phi_1^{(\beta )} \rangle \per(A^{(\alpha \beta )}  ;1|1 ) + \sum_{j_1 =2 }^N \sum_{i_1 =2 }^N\langle \phi_1^{(\alpha )} | \phi_{j_1}^{(\beta )} \rangle \langle \phi_{i_1}^{(\alpha )} | \phi_1^{(\beta )} \rangle \per(A^{(\alpha \beta )} ;1,i_1|1,j_1 ) .
\end{eqnarray}
Here $\per(A^{(\alpha \beta )}; i|j )$ denotes the permanent of the $(N-1)\times(N-1)$ minor obtained by deleting from $A^{(\alpha \beta)}$ its $i$th row and $j$th column. Similarly, $\per(A^{(\alpha \beta )}; i_1, i_2 |j_1 ,j_2 )$ denotes the permanent of the $(N-2)\times (N-2)$ minor obtained by deleting from $A^{(\alpha \beta )}$ row $i_1$, $i_2$ and column $j_1$, $j_2$.
From this expression, we can read off the operator $\hat{G}^{(\alpha \beta )} $. It is
\begin{eqnarray}
	\hat{G}^{(\alpha \beta )} &=& \hat{\mathcal{G}}(\phi_2^{(\alpha )}, \ldots, \phi_N^{(\alpha )} ; \phi_2^{(\beta )}, \ldots, \phi_N^{(\beta )} )   \nonumber \\ 
	&=&  \per(A^{(\alpha\beta  )};1|1 ) \hat{I} + \sum_{j_1 =2 }^N \sum_{i_1 =2 }^N    \per(A^{(\alpha \beta )};1,i_1|1,j_1 )  | \phi_{j_1}^{(\beta )} \rangle \langle \phi_{i_1} ^{(\alpha )}|,
\end{eqnarray}
where $\hat{I}$ is the  $L\times L $ identity matrix. Once we have constructed $\hat{F}^{(\alpha \beta )}$ and $\hat{G}^{(\alpha \beta )}$, and thus  $\hat{F}$ and $\hat{G}$ \cite{numerical}, the optimal $\phi_1$ is obtained by solving the generalized eigenvalue problem $\hat{F} \phi = \epsilon \hat{G} \phi $ \cite{linear}. It is the generalized eigenvector corresponding to the smallest generalized eigenvalue. After updating $\phi_1 \equiv  (\phi_1^{(1)}; \phi_1^{(2)} ; \ldots ; \phi_1^{(M)} )  $, we can proceed to $\phi_2 \equiv  (\phi_2^{(1)}; \phi_2^{(2)} ; \ldots ; \phi_2^{(M)} ) $ and update it similarly, and so on. In practice, for the ease of coding, we simply shift the orbitals circularly $\phi_i \rightarrow \phi_{i-1}$, and continue to update $\phi_1$. In this way, the variational energy $E $ decreases monotonically, and as it is lower bounded by the exact ground state energy, it will definitely converge. 

\subsection{Variation after projection (VAP) scheme}\label{secformulae}

A single- or multi-configurational variational wave function (\ref{mconf}), which is obtained by the algorithm above, often breaks the symmetry of the Hamiltonian. This is not surprising at all. It is even anticipated in many cases. For example, for the one-dimensional Bose-Hubbard model with the periodic boundary condition, by the Frobenius-Perron theorem, it is easily proven that the exact ground state must be positive everywhere in the real space and it must belong to the $q = 0 $ quasi-momentum subspace \cite{ed}. That is, it must be invariant under the translation (or rotation) on the lattice ring. However, suppose the particle number $N$ and the lattice size $L $ are coprime (say, $N = 10 $ and $L = 11 $), the energy-wise optimal single-configurational variational wave function almost surely will break the translation symmetry. How could a configuration of 10 wave-packets on a lattice ring of 11 sites be translationally invariant? The only configuration we can think of is to put all the particles in the $q=0$ Bloch state, but this configuration is definitely not energetically optimal when the on-site interaction is strong. 

For a few-body wave function, symmetry-breaking is generally an undesirable feature both aesthetically and practically, aesthetically because it violates rigorous rules and practically because it impairs the accuracy of predicted values of various physical quantities. It is thus necessary to restore the symmetry. A very straightforward approach is ``projection after variation'', in which the wave function is simply projected onto the desired symmetry space. This is exactly how we constructed the Slater determinant and the permanent in Eqs.~(\ref{slater}) and (\ref{permstate}). The product state $\phi_1(x_1) \ldots \phi_N(x_N )$ is neither symmetric nor anti-symmetric, but we can make it symmetric or anti-symmetric by applying the corresponding projection operator to it. 

In our current problem, suppose the symmetry group of the model is $G = \{ \hat{g} | \hat{g} \hat{H}  = \hat{H} \hat{ g}  \}$, and suppose we have obtained an energy-minimizing variational state $\Phi $ in the form of (\ref{mconf}). We can construct a wave function belonging to any irreducible representation of $G $ by the standard projection operator method \cite{elliott}. Let $D^{(\gamma  )}$ be an irreducible unitary representation of $G $ and let $d_\gamma $ be its dimension, then the operator
\begin{eqnarray}\label{projope}
	\hat{P}_{ij }^{(\gamma )} = \frac{d_\gamma}{|G|} \sum_{\hat{g}\in G }   D_{ij}^{(\gamma) *} (\hat{g}) \hat{g} , \quad\quad  1\leq i,j \leq d_\gamma, 
\end{eqnarray}
when acting on $\Phi $, yields a wave function
\begin{eqnarray}\label{projectedwf}
	\Phi_{ij}^{(\gamma)} = \frac{d_\gamma}{|G|} \sum_{\hat{g} \in G } \sum_{\alpha=1}^M   D_{ij}^{(\gamma) *} ( \hat{g})  \hat{\mathcal{S}}( \hat{g}\phi^{(\alpha)}_1, \ldots, \hat{g} \phi^{(\alpha)}_N ) ,
\end{eqnarray}
which transforms according to the $i$th row of the representation $D^{(\gamma)}$. That is, we have
\begin{eqnarray}
	\hat{g} \Phi_{ij}^{(\gamma)} = \sum_{k=1}^{d_\gamma } D_{ki}^{(\gamma )} ( \hat{g}) \Phi_{kj}^{(\gamma)} ,
\end{eqnarray}
for arbitrary group element $\hat{g} $. In particular, as for many models, the exact ground state belongs to the trivial representation of $G$, we can construct the averaged state 
\begin{eqnarray}
	\bar{\Phi } = \frac{1}{|G|}\sum_{\hat{g}\in G }   \hat{g} \Phi  = \frac{1}{|G|} \sum_{\hat{g}\in G }  \sum_{\alpha = 1 }^M \hat{\mathcal{S}}( \hat{g}\phi^{(\alpha)}_1, \ldots, \hat{g} \phi^{(\alpha)}_N )  . 
\end{eqnarray}
It is easily verified that $\hat{g} \bar{\Phi } = \bar{\Phi }$ for any $\hat{g} \in G $ by the group rearrangement theorem.

In Eq.~(\ref{projectedwf}), we have $d_\gamma $ projected wave functions (indexed by $j $) all belonging to the $i$th row of the irreducible representation of $D^{(\gamma )}$.  Generally, they are not orthogonal to each other but are linearly independent. We can seek an energy-minimizing state in the subspace spanned by them. That is, now the variational state is 
\begin{eqnarray}\label{vwf_final}
	\Phi_i^{(\gamma )} = \sum_{j=1}^{d_\gamma } w_j \Phi_{ij}^{(\gamma)}=\sum_{j=1}^{d_\gamma } w_j P_{ij}^{(\gamma)} \Phi ,
\end{eqnarray} 
where the $w$'s are complex weight coefficients to be determined. The energy of the variational state is 
\begin{eqnarray}\label{E3}
	E = \frac{\langle \Phi_i^{(\gamma)} | \hat{H} | \Phi_i^{(\gamma)} \rangle }{\langle \Phi_i^{(\gamma)} | \Phi_i^{(\gamma)} \rangle } = \frac{\sum_{j,k} w_j^*\langle \Phi | 	\hat{P}_{ij }^{(\gamma ) \dagger } \hat{H} 	\hat{P}_{ik }^{(\gamma )}  | \Phi \rangle  w_k  }{\sum_{j,k} w_j^* \langle \Phi | 	\hat{P}_{ij }^{(\gamma ) \dagger } 	\hat{P}_{ik }^{(\gamma )}  | \Phi \rangle w_k  } = \frac{\sum_{j,k} w_j^* \langle \Phi | 	 \hat{H} 	\hat{P}_{jk }^{(\gamma )}  | \Phi \rangle w_k }{\sum_{j,k} w_j^* \langle \Phi | 		\hat{P}_{jk }^{(\gamma )}  | \Phi \rangle  w_k }. 
\end{eqnarray}
Here in the third equality, we have used the fact that $	\hat{P}_{ij }^{(\gamma ) \dagger } = 	\hat{P}_{ji  }^{(\gamma ) }$, $ \hat{P}_{mn   }^{(\gamma )} \hat{P}_{ij   }^{(\gamma ) }= \hat{P}_{mj   }^{(\gamma ) }\delta_{ni }$, and $ [\hat{P}_{ij   }^{(\gamma ) }, \hat{H} ]=0$. Note that the last expression of the energy is independent of $i $. This means that the states $  \Phi_i^{(\gamma)} $ ($1\leq  i \leq d_\gamma $) are of the same norm and degenerate in energy. Again, we recognize quadratic forms of the weight vector $w = (w_1, w_2, \ldots, w_{d_\gamma })^T$ in the numerator and denominator of the last expression, and the optimal weight vector is obtained by solving the generalized eigenvalue problem $\mathcal{N} w = E \mathcal{D} w $, with $\mathcal{N}_{jk } = \langle \Phi | 	 \hat{H} 	\hat{P}_{jk }^{(\gamma )}  | \Phi \rangle $ and $\mathcal{D}_{jk }= \langle \Phi | \hat{P}_{jk }^{(\gamma )}  | \Phi \rangle$.

The ``projection after variation'' (PAV) approach above, while simple, has several drawbacks. First, it is not optimal, as the orbitals $\phi^{(\alpha )}_i $ ($1\leq \alpha \leq M $, $1\leq i \leq N $) are obtained by minimizing the energy of $\Phi $ in (\ref{mconf}), instead of that of the symmetry-adapted state  $ \Phi_i^{(\gamma)}  $ in (\ref{vwf_final}). It is desirable to optimize the orbitals with the latter being the variational wave function.  Second, it often happens that the pre-projection state $\Phi$ belongs to some definite symmetry sector other than $\gamma$, and thus the projected state  $ \Phi_i^{(\gamma)}  $ vanishes identically, rendering the $\gamma$ sector inaccessible from $\Phi $. Or, it could be that the component of $\Phi $ in the $\gamma $ sector is nonzero but negligible, then the projected state  $ \Phi_i^{(\gamma)}  $ is unlikely to be a good variational state. Third, in the PAV approach, all projected states originate from the same state $\Phi $. This implicitly assumes that the lowest eigenstates in all symmetry sectors share the same internal structure, which is rigid to external motion. Such a stringent requirement is generally satisfied only when $\Phi $ strongly violates symmetry \cite{ring}. 

The ``variation after projection'' (VAP) approach avoids all these difficulties. In this approach, we first construct the symmetry-adapted variational wave function $ \Phi_i^{(\gamma)}  $ in (\ref{vwf_final}), and then try to minimize its energy given in (\ref{E3}). The problem of the VAP approach, in comparison of the PAV approach, is that it is costlier, as now we have to repeat the variation for each symmetry sector again; but more importantly, it is generally more complicated and much harder to implement \cite{ring, symmetry1, symmetry2}. Fortunately, it is perfectly compatible with our formalism in the proceeding subsection and does not cause any difficulty. We note that for fixed weight vector $w$, we can again introduce the vector $\phi_1 =  (\phi_1^{(1)}; \phi_1^{(2)} ; \ldots ; \phi_1^{(M)} ) $, and the $LM\times LM $ block matrices $\hat{F} \equiv  (\hat{F}^{(\alpha \beta )})_{1\leq \alpha, \beta \leq M} $ and $\hat{G} \equiv  (\hat{G}^{(\alpha \beta )})_{1\leq \alpha, \beta \leq M }$, to rewrite the ratio (\ref{E3}) in the form of (\ref{ratio2}). The expressions of $\hat{F}^{(\alpha \beta )} $ and $\hat{G}^{(\alpha \beta )} $ are now 
\begin{subequations}
\begin{eqnarray}
		\hat{F}^{(\alpha \beta )} &=& \frac{d_\gamma}{|G|} \sum_{\hat{g}\in G } \sum_{j,k=1}^{d_\gamma} w_j^* D_{jk}^{(\gamma) *} (\hat{g} ) w_k \hat{\mathcal{F}}(\phi_{2\leq i \leq N}^{(\alpha )} ; \hat{g}\phi_{2\leq i \leq N}^{(\beta )} ) \hat{g} , \\ 
	\hat{G}^{(\alpha \beta )} &=& \frac{d_\gamma}{|G|} \sum_{\hat{g}\in G }  \sum_{j,k=1}^{d_\gamma} w_j^* D_{jk}^{(\gamma) *} (\hat{g} ) w_k \hat{\mathcal{G}}(\phi_{2\leq i \leq N}^{(\alpha )} ; \hat{g} \phi_{2\leq i \leq N}^{(\beta )} ) \hat{g}  . 
\end{eqnarray}
\end{subequations}
We see that the codes for constructing the matrices $\hat{\mathcal{F}}$ and $\hat{\mathcal{G}}$ can be reused here, and the complexity of constructing the operators $\hat{F}$ and $\hat{G}$ just increases linearly with the order of the group $G$.

The update strategy now goes as follows. First, with the weight vector $w$ fixed, we update $\phi_1 $, and then with $\phi_1$ fixed (and thus $\Phi$ fixed), we update the weight vector $w$.  Afterwards, we make the circular shift $\phi_i \rightarrow \phi_{i-1 }$ and repeat the procedure before. 

Here some technical remarks are in order. First, as preparing the single-particle operators
\begin{eqnarray*}
	\hat{\mathcal{F}}(\phi_{2\leq i \leq N}^{(\alpha )} ; \hat{g} \phi_{2\leq i \leq N}^{(\beta )} ) \hat{g} \quad \text{ and  } \quad \hat{\mathcal{G}}(\phi_{2\leq i \leq N}^{(\alpha )} ; \hat{g} \phi_{2\leq i \leq N}^{(\beta )} ) \hat{g}
\end{eqnarray*}
is the most time-consuming part of the optimization procedure, it is desirable to reduce the effort as much as possible. We note that a reduction by about one half is possible for large $M$ (from $M^2$ to $M(M+1 )/2$ actually), as we have 
\begin{subequations}\label{identity}
\begin{eqnarray}
	\hat{\mathcal{F}}(\phi_{2\leq i \leq N}^{(\alpha )} ; \hat{g}\phi_{2\leq i \leq N}^{(\beta )} ) \hat{g} &=& [\hat{\mathcal{F}}(\phi_{2\leq i \leq N}^{(\beta )} ; \hat{g}^{-1}\phi_{2\leq i \leq N}^{(\alpha )} ) \hat{g}^{-1} ]^\dagger, \\
	\hat{\mathcal{G}}(\phi_{2\leq i \leq N}^{(\alpha )} ; \hat{g}\phi_{2\leq i \leq N}^{(\beta )} ) \hat{g} &=& [\hat{\mathcal{G}}(\phi_{2\leq i \leq N}^{(\beta )} ; \hat{g}^{-1}\phi_{2\leq i \leq N}^{(\alpha )} ) \hat{g}^{-1} ]^\dagger. 
\end{eqnarray}
\end{subequations}
The proof is simple as 
\begin{eqnarray}
	\langle \phi_1^{(\alpha) }| \hat{\mathcal{F}}(\phi_{2\leq i \leq N}^{(\alpha )} ; \hat{g} \phi_{2\leq i \leq N}^{(\beta )} ) \hat{g}| \phi_1^{(\beta)}\rangle &=& \langle \Phi^{(\alpha)} | \hat{H} \hat{g}| \Phi^{(\beta)} \rangle \nonumber \\
	&=& \langle \Phi^{(\beta)} |\hat{g}^{-1} \hat{H} | \Phi^{(\alpha)} \rangle^* \nonumber \\
		&=& \langle  \Phi^{(\beta)} | \hat{H} |g^{-1}  \Phi^{(\alpha)} \rangle^* \nonumber \\
		&=& \langle \phi_1^{(\beta) }| \hat{\mathcal{F}}(\phi_{2\leq i \leq N}^{(\beta )} ; \hat{g}^{-1}\phi_{2\leq i \leq N}^{(\alpha )} ) \hat{g}^{-1}| \phi_1^{(\alpha)}\rangle^*.
\end{eqnarray}
This equation holds for arbitrary single-particle orbitals $\phi_1^{(  \alpha )} $ and $\phi_1^{(\beta ) } $, which proves (\ref{identity}a). Similarly, (\ref{identity}b) can be proven. These relations allow us to prepare only the operators with $\alpha \leq \beta $. Second, although the variational wave function in (\ref{vwf_final}), with the $w$'s treated as variational parameters, is the most general and optimal for accuracy, in practice, we often just fix the $w$'s to the particular value $w_j = \delta_{ij }$. That is, we simply take $\Phi_i^{(\gamma )}  = \hat{P}^{(\gamma)}_{ii} \Phi $. While this approach compromises accuracy, it offers several advantages. On the one hand, fixing the $w$'s simplifies the coding; on the other hand, the matrix element $D_{ii}^{(\gamma)} (\hat{g})$ often vanishes for many group elements $\hat{g}$, reducing the need to prepare the $\hat{\mathcal{F}}$ and $\hat{\mathcal{G}}$ operators involving these elements. Consequently, this saves CPU time while maintaining the same number of configurations. The lost accuracy can be compensated by increasing the number of configurations.

Finally, we would like to mention that the projection operator method is also used in the exact diagonalization calculation of the lowest eigenstate belonging to the $i$th row of the representation $D^{(\gamma)}$. Consider the hermitian  operator
\begin{eqnarray}
  \tilde{H } &=& \hat{H} \hat{P}_{ii}^{(\gamma )} + \lambda (1- \hat{P}_{ii}^{(\gamma )}) ,
\end{eqnarray}
where $\lambda$ is an adjustable real parameter. By construction, in the subspace of $\hat{P}_{ii}^{(\gamma )}$, $\tilde{H }$ coincides with $ \hat{H} $; while in the orthogonal complement of $ \hat{P}_{ii}^{(\gamma )}$, $\tilde{H }$ is a constant of $\lambda $. Therefore, as long as $\lambda $ is larger than the minimal eigenvalue of $\hat{H} $ in the subspace of $ \hat{P}_{ii}^{(\gamma )}$, the ground state of $\tilde{H }$ is the desired eigenstate of $ \hat{H} $, and we
can use the Lanczos method to solve it \cite{linearalgebra}. In this paper, in all case studies below, we find that we can simply take $\lambda = 0 $ and thus $\tilde{H } =
 \hat{H} \hat{P}_{ii}^{(\gamma )} $. As we use the Lanczos method, we do not need to construct $\tilde{H }$ or even $ \hat{P}_{ii}^{(\gamma )}$ explicitly. We just need to prepare the sparse matrices of $ \hat{H} $ and some generators of the group $G$. That is then sufficient for realizing the matrix-vector multiplication $v\rightarrow \tilde{H } v = \hat{H} (\hat{P}_{ii}^{(\gamma )} v )$, which is all the Lanczos method needs. 

Unlike other common approaches \cite{ed,edbook}, this projection operator strategy works in the full many-body Hilbert space and always with the real-space basis. It does not bother to construct a basis for the subspace $  \hat{P}_{ii}^{(\gamma )}$. Although this is not necessarily economical in memory, it greatly simplifies the implementation. 

\section{Case studies}\label{secbhm}

We now proceed to apply the numerical schemes to various models with different symmetries. 

\subsection{$\mathbb{Z}_2$ symmetry}

\begin{figure*}[tb]
	\centering
	\includegraphics[width= 0.32\textwidth ]{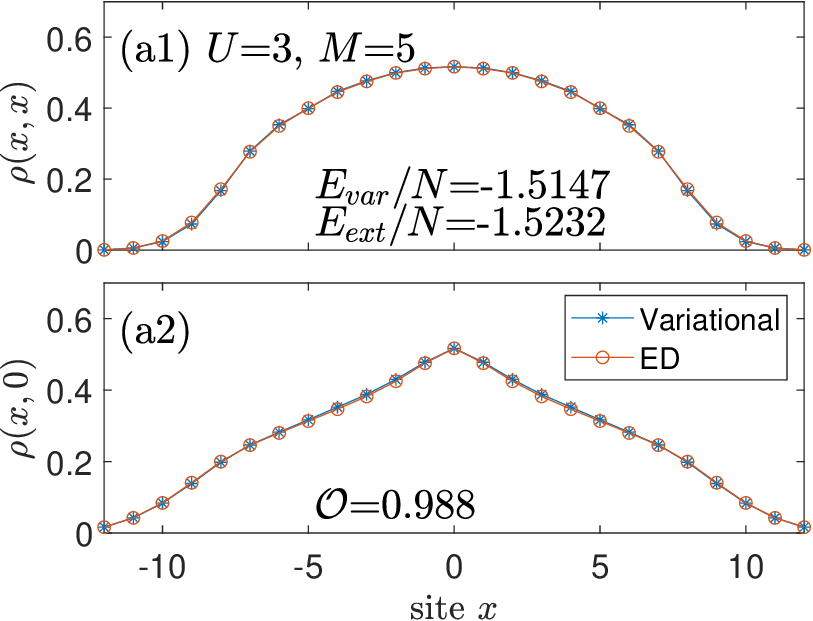}
	\includegraphics[width= 0.32\textwidth ]{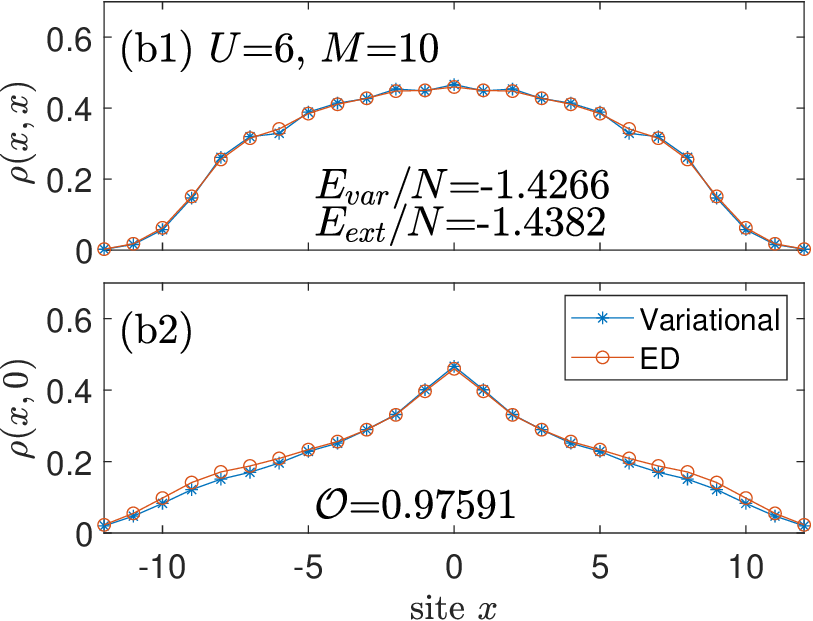}
	\includegraphics[width= 0.32\textwidth ]{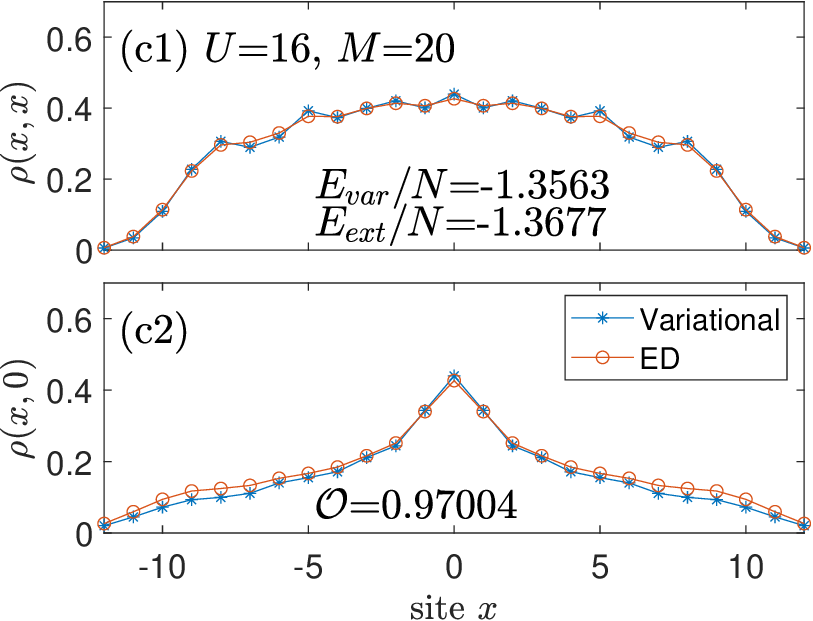}
	\caption{(Color online) Diagonal (upper panels) and off-diagonal (lower panels) elements of the one-particle reduced density matrix $\rho(x, y ) = \langle \hat{a}_y^\dagger \hat{a}_x \rangle $ of the lowest even-parity eigenstate of the one-dimensional Bose-Hubbard model with the open boundary condition [see Eq.~(\ref{hamobc})]. Results obtained by the symmetry-adapted permanent variational wave function $\Phi^{(+)}$ in (\ref{phiplus}) are compared with those by the exact diagonalization (ED) method. In each column, the value of the on-site interaction strength $U $ and the the number of independent configurations $M$ are shown at the left upper corner. The variational energy $E_{var}$, the exact energy $E_{ext}$, and the overlap $\mathcal{O}$ between the variational wave function and the exact one  are also displayed. The common parameters are $N = 7$, $L= 25$, and $k = 0.02$. The dimension of the many-body Hilbert space is $2\,629\,575$.  }
	\label{fig_z2_p1}
\end{figure*}

\begin{figure*}[tb]
	\centering
	\includegraphics[width= 0.32\textwidth ]{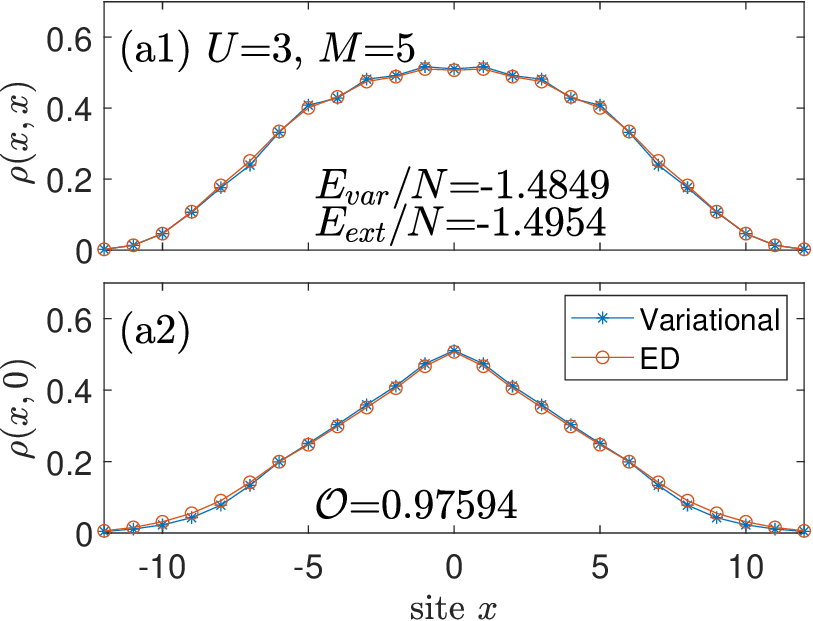}
	\includegraphics[width= 0.32\textwidth ]{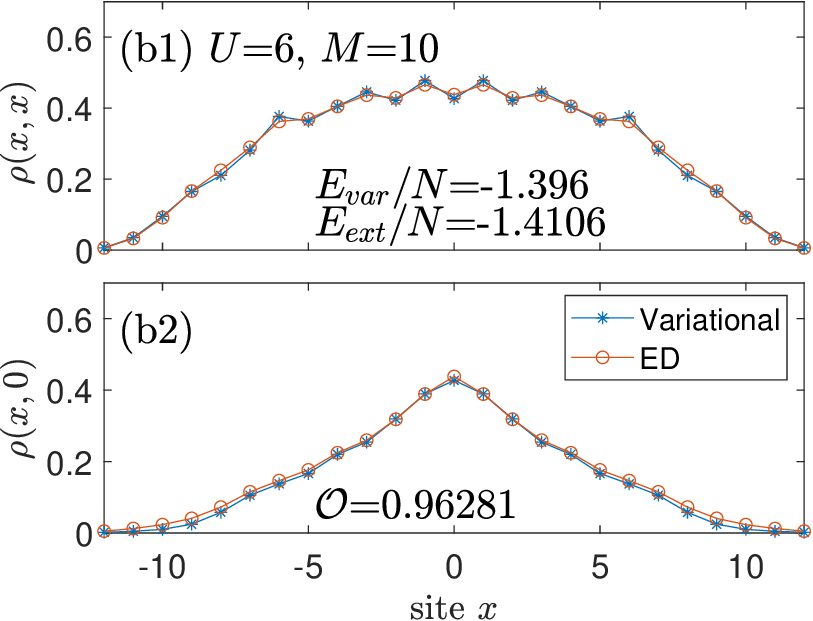}
	\includegraphics[width= 0.32\textwidth ]{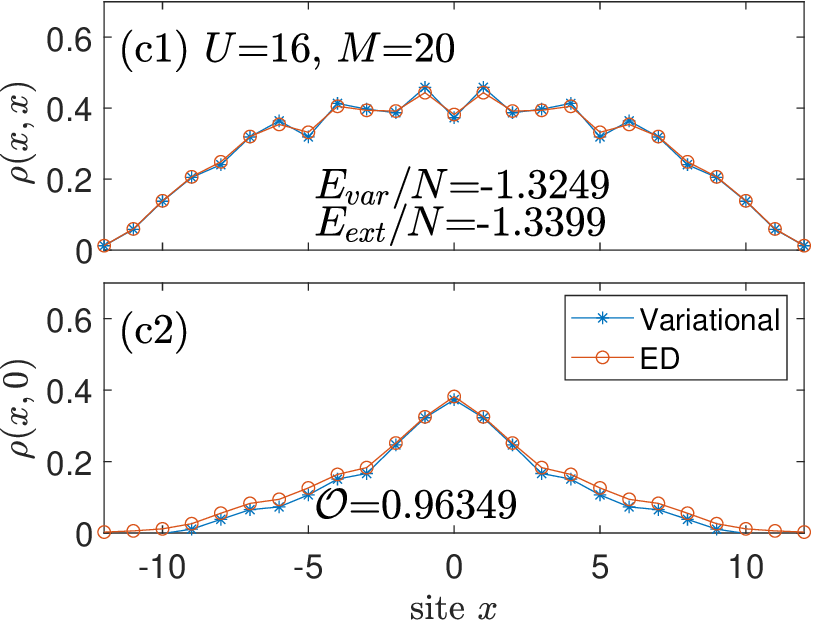}
	\caption{(Color online) Same as Fig.~\ref{fig_z2_p1} but for the lowest odd-parity eigenstate of (\ref{hamobc}).  }
	\label{fig_z2_pm1}
\end{figure*}

\begin{figure}[tb]
	\centering
	\includegraphics[width= 0.45\textwidth ]{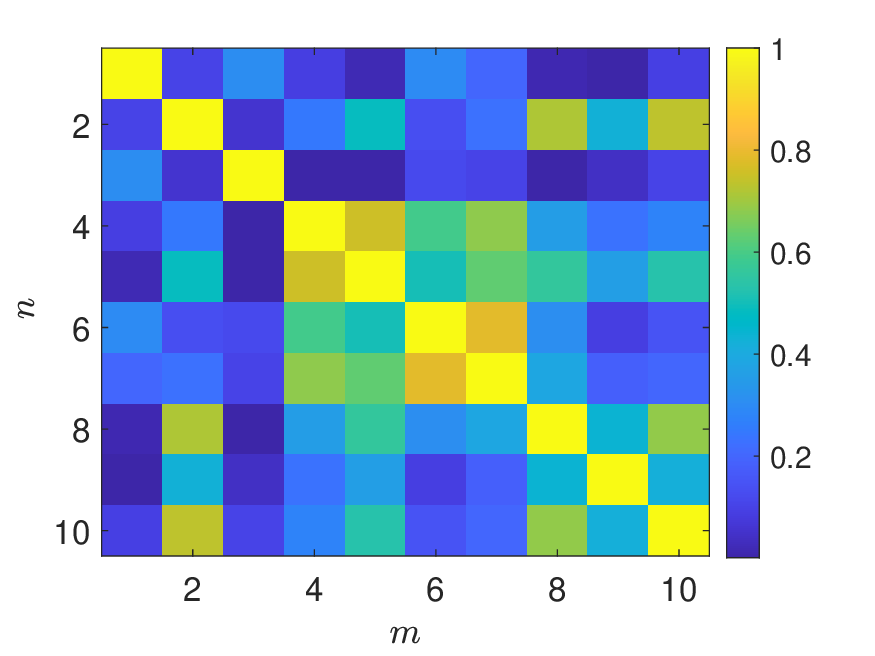}
	\caption{(Color online) Overlaps between the pre-projection states [see Eq.~(\ref{gram})]. The setting is as in Fig.~\ref{fig_z2_p1}(a1). We have ran the code ten times to generate ten projected variational state $\Phi^{(+)}_m $ [see Eq.~(\ref{phiplus})], $1\leq m \leq 10 $. Their overlap with the exact ground state $|GS\rangle $ is at least 0.9841, which means that they are very close to each other. However,   as demonstrated by the smallness of the off-diagonal elements here, the pre-projection states $\Phi_m $ are not necessarily close to each other.  }
	\label{fig_gram}
\end{figure}

We start from the simplest nontrivial group, namely, the $\mathbb{Z}_2$ group. This symmetry is realized by a one-dimensional Bose-Hubbard model with the open boundary condition and in the presence of a harmonic trap. The Hamiltonian is ($L =2L_1 +1 $)
 \begin{eqnarray}\label{hamobc}
	\hat{H} &=& - \sum_{x=-L_1}^{L_1-1} (\hat{a}_x^\dagger \hat{a}_{x+1} + \hat{a}_{x+1}^\dagger \hat{a}_{x}) + \frac{k}{2}\sum_{x=-L_1}^{L_1} x^2  \hat{a}_x^\dagger \hat{a}_x   + \frac{U }{2} \sum_{x=-L_1}^{L_1}  \hat{a}_x^\dagger \hat{a}_x^\dagger \hat{a}_x \hat{a}_x .\quad
\end{eqnarray}
The symmetry group is $\mathbb{Z}_2 = \{ \hat{I}, \hat{P} \}$, where $\hat{P}$ is the reflection operator. Its action on a single-particle orbital $\phi(x) $ is $(\hat{P} \phi) (x) = \phi(-x)$, and the induced action on a many-particle state is $(\hat{P}f)(x_1, \ldots, x_N) = f(-x_1, \ldots, -x_N )$. The eigenstates are classified into even- and odd-parity ones according to the eigenvalues of $\hat{P}$. The associated projection operators are 
$ \hat{P}^{(\pm)} = (\hat{I}\pm \hat{P})/2 $. The even- or odd-parity variational states are then of the form 
\begin{eqnarray}\label{phiplus}
	\Phi^{(\pm)}  =\frac{1}{2}(\Phi \pm \hat{P} \Phi ) = \frac{1}{2}\sum_{\alpha =1}^M \left [  \hat{\mathcal{S}}( \phi^{(\alpha)}_1, \ldots, \phi^{(\alpha)}_N )\pm  \hat{\mathcal{S}}( \hat{P}\phi^{(\alpha)}_1, \ldots, \hat{P} \phi^{(\alpha)}_N )  \right ]. 
\end{eqnarray}

In Fig.~\ref{fig_z2_p1}, we take the even-parity state $\Phi^{(+)} $ to target the ground state $|GS \rangle $ of the Hamiltonian (\ref{hamobc}), which is also of even-parity.  We have chosen a system of $N=7 $ particles and $L = 25 $ sites. The  Hilbert space dimension is $\mathcal{D} = 2\,629\,575 $, which is large enough to provide a stringent test of the expressibility of the symmetry-adapted permanent variational wave function, yet not so large as to preclude exact diagonalization. As the on-site repulsion $U $ increases from $3$ to $6$ and then to $16$, the system becomes increasingly fragmented, requiring more and more configurations to achieve a good approximation of the exact ground state. But anyway, for $U = 3, 6, 16$, this objective can be achieved by taking $M=5,10,20$ independent configurations, respectively.  We see that the relative error in energy is less than $0.85\%$, and the overlap between the variational state and the exact ground state, $\mathcal{O}= |\langle GS | \Phi^{(+)}\rangle |^2$, is at least 0.970. The high overlap suggests that the variational wave function can produce many physical quantities accurately. Indeed, as shown in Fig.~\ref*{fig_z2_p1}, the density distribution $\rho(x,x ) = \langle \hat{a}_x^\dagger \hat{a}_x \rangle $ and the correlator $\rho(x,0)= \langle \hat{a}_0^\dagger \hat{a}_x \rangle $ predicted by $\Phi^+ $ agree very well with the exact values. Moreover, because of the built-in symmetry of $\Phi^{(+)} $, these functions are now strictly even in $x$. Symmetry is restored! 

In Fig.~\ref{fig_z2_pm1}, we examine the lowest eigenstate in the \emph{odd-parity} sector. The variational state is now $\Phi^{(-)} $. Once again, we see that the variational wave function can achieve high overlap with the targeted exact state and in turn, deliver the energy and other physical quantities like the one-body reduced density matrix accurately. 

We have thus seen that the symmetry-adapted variational wave function $\Phi^{(\pm)} $ can approximate the target state to high accuracy. However, since it results from a projection, which is not invertible, one might ask: does the pre-projection state $\Phi $ converge to a definite state too? Or, is the state $\Phi $, composed of the independent permanent states (\ref{mconf}), unique? The answer is no. To demonstrate this point, we have taken the case of Fig.~\ref{fig_z2_p1}(a), and have ran the code ten times to generate ten $\Phi^{(+)}$, $\{\Phi_m^{(+)} = (\Phi_m + \hat{P} \Phi_m )/2, 1\leq  m \leq 10  \}$. Each $\Phi_m^{(+)}$ is a highly accurate approximation of $|GS \rangle $---the overlap is at least 0.9841. To investigate the closeness of the pre-projection states $\Phi_m$, we consider the element-wise squared Gram matrix 
\begin{eqnarray}\label{gram}
	A_{mn } = \frac{|\langle \Phi_m|\Phi_n \rangle  |^2}{\langle \Phi_m |\Phi_m \rangle \langle \Phi_n | \Phi_n \rangle } . 
\end{eqnarray} 
In Fig.~\ref{fig_gram}, we see that many of the off-diagonal matrix elements are significantly smaller than unity. This means that the pre-projection states are substantially different from each other, although they all project close to the same state. Therefore, the pre-projection state $\Phi $ does not need to converge to a ``point'' to deliver a high-quality approximation of the target state, but only needs to converge towards a higher-dimensional ``manifold''. This has important implications on the performance of the algorithm. In general, in a multi-variable optimization problem, one is often plagued by the problem of being trapped in local minima. However, according to our experience, this is not a problem, at least not a severe one, in our case. Our experience is that,  regardless of the initial states, as long as we iterate sufficiently many times, the final post-projection states we obtain will be close to each other. Therefore, in this paper, we run each simulation only once. We stop the iteration once the relative improvement in energy, compared to its previous value, falls below $10^{-8 }$. 

\subsection{$\text{D}_L$ symmetry}\label{dlsym}

A paradigmatic model realizing the $\text{D}_L $ symmetry is the Bose-Hubbard model on a one-dimensional lattice of $L $ sites and  with the periodic boundary condition. The Hamiltonian is 
 \begin{eqnarray}\label{bhm}
	\hat{H} = - \sum_{x=1}^{L} (\hat{a}_x^\dagger \hat{a}_{x+1} + \hat{a}_{x+1}^\dagger \hat{a}_{x})   + \frac{U }{2} \sum_{x=1}^{L}  \hat{a}_x^\dagger \hat{a}_x^\dagger \hat{a}_x \hat{a}_x .\quad
\end{eqnarray}
Imaging that the lattice is closed as a circle (because of the periodic boundary condition) and the lattice sites are arranged uniformly on the circle, we recognize a regular $L$-sided polygon and see that the symmetry group of the model is the dihedral group $\text{D}_L$. This group is generated by the translation operator $\hat{T}$ and the reflection operator $\hat{S}$, which act on a single-particle wave function $\phi (x)$ as $(\hat{T}\phi )(x) = \phi(x-1)$ and $(\hat{S}\phi )(x) = \phi(- x)$, respectively. The induced actions on a many-particle wave function are obvious and therefore are omitted here for brevity. Explicitly, the group $\text{D}_L $ is 
\begin{eqnarray}\label{DL}
	\text{D}_L = \{ \hat{I}, \hat{T}, \ldots, \hat{T}^{L-1}, \hat{S}, \hat{S} \hat{T} ,\ldots, \hat{S} \hat{T}^{L-1} \} . 
\end{eqnarray}
We have $\hat{T}^L = \hat{S}^2 = \hat{I}$ and $\hat{S} \hat{T} \hat{S} = \hat{T}^{-1}$. 

We would like to classify the eigenstates of $H $ according to the irreducible representations of $\text{D}_L $. To this end, we start from the cyclic  group of translations, $\text{C}_L = \{ \hat{I}, \hat{T}, \ldots, \hat{T}^{L-1}   \}$,  which is a normal subgroup of $\text{D}_L $ with index 2. As an abelian group, all irreducible representations of $\text{C}_L $ are one-dimensional. They are labelled by an integer $q$ modulo $L$. A state belongs to the $q$th representation if it is an eigenstate of $\hat{T}$ with eigenvalue $e^{i 2 \pi q /L }$. We call $q$ the quasi-momentum. Now suppose $\Psi $ belongs to the $q$th representation or the $q$-sector, then the reflected state $\hat{S} \Psi $ belongs to the $(-q)$-sector, as $\hat{T} \hat{S} \Psi = \hat{S} \hat{S }\hat{T} \hat{S} \Psi = \hat{S} \hat{T}^{-1}\Psi = e^{-i 2 \pi q /L }\hat{S}\Psi  $. If $q \not\equiv -q \pmod{L}$, the two sectors are disjoint and  one can verify that $\{\Psi, \hat{S}\Psi \}$ form a two-dimensional irreducible representation of $\text{D}_L$. Explicitly, 
\begin{eqnarray}
	D^{(q)}(\hat{T}^n ) = \begin{pmatrix}
		e^{i 2 \pi q n /L} & 0 \\ 0  & e^{-i 2 \pi q n /L }
	\end{pmatrix} , \quad D^{(q)} (\hat{S}\hat{T}^n ) =  \begin{pmatrix}
	0 & e^{-i 2 \pi q n /L}  \\ e^{i 2 \pi q n /L } & 0 
	\end{pmatrix} . 
\end{eqnarray}
However, if $q \equiv -q \pmod{L}$ or $2q \equiv 0 \pmod{L }$, the two sectors are identical, and we have that $\hat{S}$ acts on the  $q$-sector as an involution. The $q$-sector can then be further divided into an even-subsector and an odd-subsector according to the eigenvalue $s$ of $\hat{S}$. In other words, the original $q$th representation of $\text{C}_L $ splits into two one-dimensional representations of $\text{D}_L $, 
\begin{eqnarray}
	D^{(q, s)}(\hat{T}^n) =e^{i 2 \pi q n /L }, \quad 	D^{(q, s)}(\hat{S} \hat{T}^n) = s e^{i 2 \pi q n /L} ,
\end{eqnarray}
where $s = \pm 1 $. We have thus constructed all the irreducible representations of $\text{D}_L $. For even $L$, we get four one-dimensional (with $q= 0, L/2$) and $(L-2)/2$ (with $1\leq q \leq L/2 -1$) two-dimensional representations; while for odd $L$, we get two one-dimensional (with $q = 0 $) and $(L-1)/2$ (with $1\leq q \leq (L-1)/2$) two-dimensional representations. 

We can then use the projection operators and the projected variational wave functions to target the lowest eigenstate in the desired subspace. Specifically, for $q\neq 0 ,L/2$, we target the lowest eigenstate in the $q$-sector with the variational state $\Phi_1^{(q)} = \hat{P}^{(q)}_{11} \Phi $, or more explicitly, 
\begin{eqnarray}\label{projectedwf2}
	\Phi_1^{(q)} = \frac{1}{L} \sum_{n=0}^{L-1} \sum_{\alpha=1}^M   e^{-i 2\pi q n /L }  \hat{\mathcal{S}}( \hat{T}^n \phi^{(\alpha)}_1, \ldots, \hat{T}^n \phi^{(\alpha)}_N ). 
\end{eqnarray}
This corresponds to the $w$-fixing scheme in (\ref{vwf_final}) with $i=1$ and $w_j = \delta_{j1}$. 
For $q=0$ or $L/2$, the lowest even ($s=1$) or odd ($s= -1$) states can be targeted with 
\begin{eqnarray}\label{projectedwf2}
	\Phi^{(q,s)} = \frac{1}{L} \sum_{n=0}^{L-1} \sum_{\alpha=1}^M    e^{-i 2\pi q n /L }\left[  \hat{\mathcal{S}}( \hat{T}^n \phi^{(\alpha)}_1, \ldots, \hat{T}^n \phi^{(\alpha)}_N ) + s \hat{\mathcal{S}}(\hat{S}\hat{T}^n \phi^{(\alpha)}_1, \ldots, \hat{S}\hat{T}^n \phi^{(\alpha)}_N) \right]. 
\end{eqnarray}
In this way, we can estimate the lowest eigenvalue in each $q$-sector, and obtain the so-called Yrast line. Below, we consider the attractive and repulsive cases separately, as the two cases are quite different.

\subsubsection{Attractive case: $U< 0 $ (Few-body bound states)}
\begin{figure*}[tb]
	\centering
	\includegraphics[width= 0.48\textwidth ]{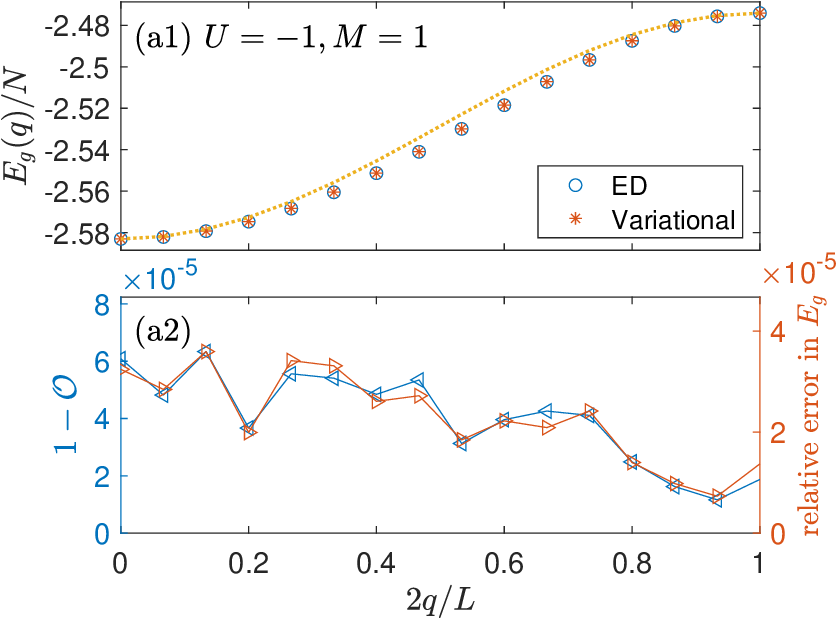}
	\includegraphics[width= 0.48\textwidth ]{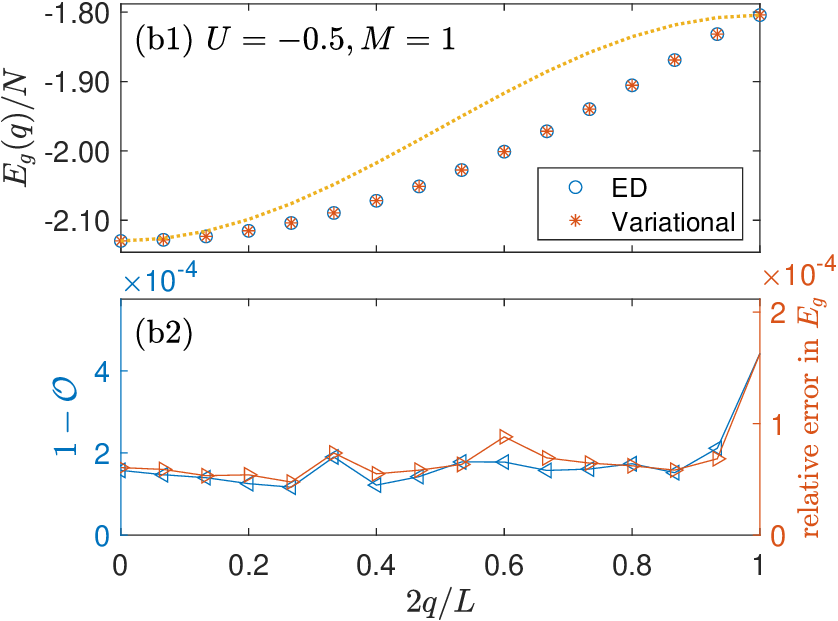}
	\caption{(Color online) Upper panels: Dispersion curve of the suspected few-body bound state of an attractive Bose-Hubbard model (\ref{bhm}). For each quasi-momentum $0\leq q\leq \floor{L/2}$, we calculate the energy of the lowest eigenstate in the $q$-sector with both exact diagonalization (ED) and the projected variational wave function with a single ($M=1$) configuration. The dotted lines are guide for the eye. They display functions of the form $A + B \cos (2\pi q /L )$. Lower panels: Overlap $\mathcal{O}$ between the exact state and the projected variational wave function (left y-axis), and the relative error in energy (right y-axis). The common parameters are $N = 5$ and $L = 30$.}
	\label{fig_fewbody}
\end{figure*}

In the attractive case, we have the picture that the bosons first bind into a molecule, which then delocalizes across the lattice with a well-defined quasi-momentum $q$. We also expect that the stronger the attraction is, the tighter the bosons are bound, the more rigid the molecule is, and the weaker its hopping amplitude on the lattice is, but the more closer its dispersion curve is to a cosine function. This development of a cosine-like dispersion curve is reminiscent to the formation of a rotational band of a deformed nucleus. 

In Fig.~\ref{fig_fewbody}, we study the Yrast line of a few-body ($N=5$) attractively interacting Bose-Hubbard model both by exact diagonalization and by the variational method. 
It can be proven that for $q=0 $ or $L/2$, the lowest eigenstate belongs to the even ($s = 1$) subsector. First, if we work in the real-space Fock state basis, we can easily see that the matrix of the Hamiltonian is irreducible and has the property that all its nonzero off-diagonal elements are negative. We can then apply the Frobenius-Perron theorem to show that the global ground state of the Hamiltonian is non-degenerate and positive everywhere. It follows that it must be an eigenstate of both the translation operator $\hat{T}$ and the reflection operator $\hat{S}$ with eigenvalue $+1$, which means it is of quasi-momentum $q=0$ and of even-parity under reflection. Second, if we work in the momentum-space Fock state basis, again we have that in each $q$-sector, the matrix of the Hamiltonian is irreducible and all its nonzero off-diagonal elements are negative (proportional to $U $). We can then use the Frobenius-Perron theorem again and obtain the result that the lowest eigenstate in each $q$-sector is positive everywhere in the momentum-space Fock basis. It follows easily that the state is even under reflection if $q=0 $ or $L/2$.  

We see that the projected variational wave function gets both the eigenstate and the eigenenergy very accurately. The relative error in energy and the deficit in overlap ($1 -\mathcal{O}$) are both on the order of $10^{-4}$ in the whole Brillouin zone. Note that we have only a single independent configuration ($M=1$), so the number of parameters of the variational wave function is about $N*L = 150$, while the dimension of the full Hilbert space is $278\,256$.

Comparing Fig.~\ref{fig_fewbody}(a) with \ref{fig_fewbody}(b), we see that indeed, as the attraction increases (here from $U=-0.5$ to $U= -1$), the width of the Yrast line shrinks and its shape approaches a cosine function. This means that in the strong coupling limit ($|U|\rightarrow \infty $), the internal motion of the molecule can be ignored, allowing it to be treated as a structureless point-particle hopping on the lattice. In contrast, in the weak coupling limit ($|U|\rightarrow 0 $), the internal motion of the molecule remains coupled to its center-of-mass motion, making the point-particle approximation inadequate. 

Finally, we note that solving the Yrast line for a few-body system, or determining whether such a system can form a bound state, is generally challenging both analytically and numerically \cite{mattis,valiente, bic1, bic2, bic3}. The high accuracy of the projected variational wave function demonstrated here suggests that our approach could be valuable in tackling these types of problems.

\subsubsection{A digression: Comparison of projection after variation (PAV)  and variation after projection (VAP)}

\begin{figure*}[tb]
	\centering
	\includegraphics[width= 0.48\textwidth ]{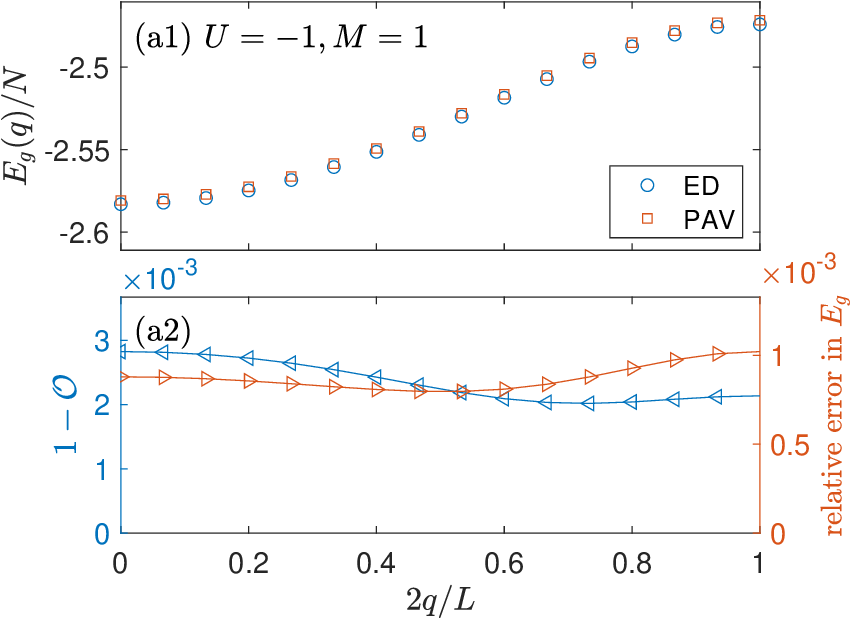}
	\includegraphics[width= 0.48\textwidth ]{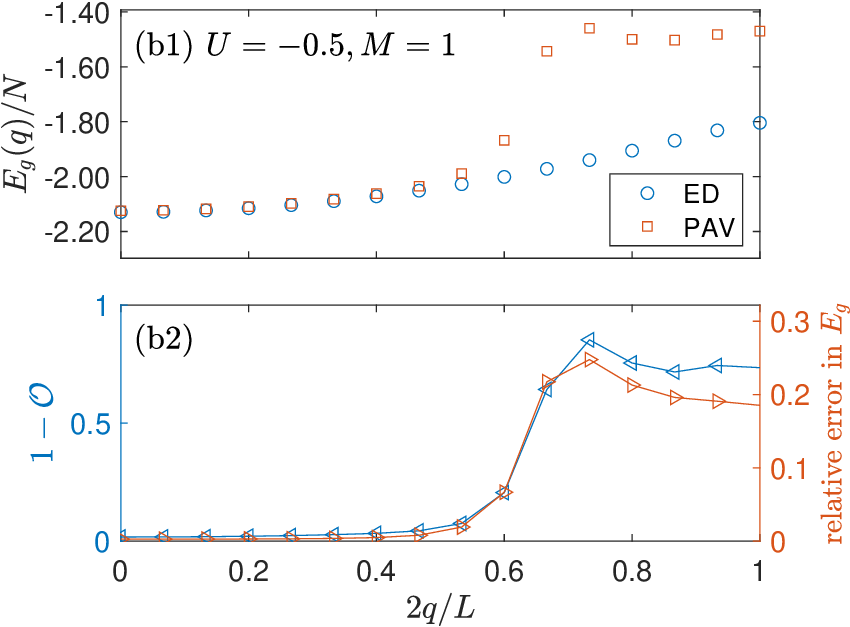}
	\caption{(Color online) This figure is to be compared with Fig.~\ref{fig_fewbody} panel by panel.  Here we study the same systems with the same parameters and calculate the same quantities. The difference is that here the variational results are obtained with the projection after variation (PAV) method, while in Fig.~\ref{fig_fewbody} the results are with the variation after projection (VAP) method. }
	\label{fig_fewbody_vbp}
\end{figure*}

\begin{figure*}[tb]
	\centering
	\includegraphics[width= 0.48\textwidth ]{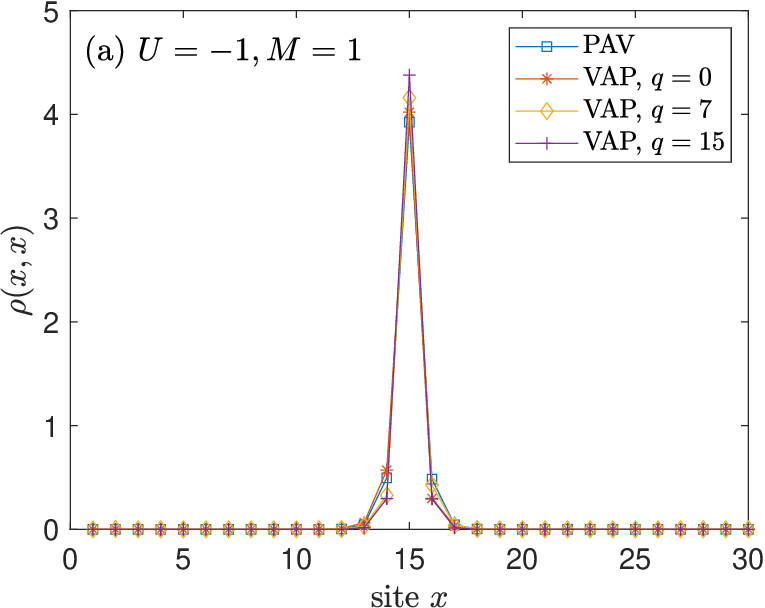}
	\includegraphics[width= 0.48\textwidth ]{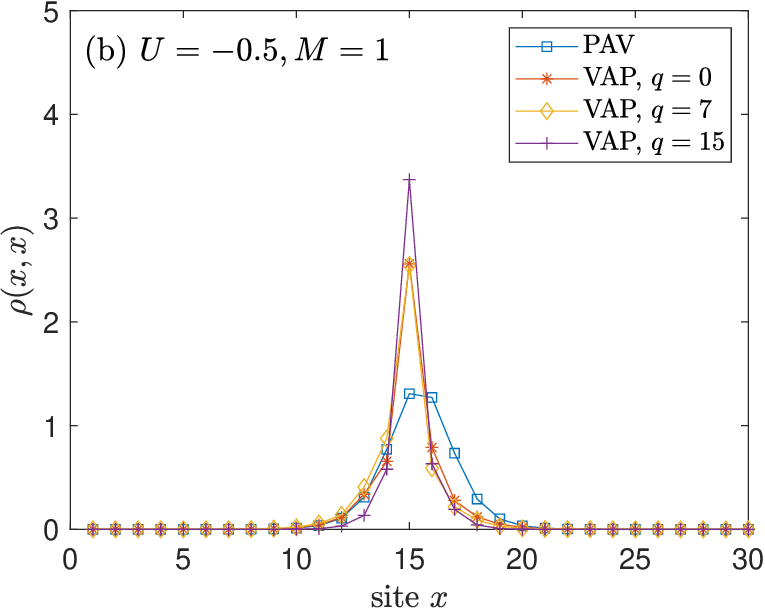}
	\caption{(Color online) Density distribution $\rho(x,x) \equiv  \langle\hat{a}_x^\dagger \hat{a}_x \rangle $  of the pre-projection state $\Phi $ in the VAP and PAV calculations in Fig.~\ref{fig_fewbody} and Fig.~\ref{fig_fewbody_vbp}. Note that due to the translational invariance, each $\Phi$ can be shifted freely across the lattice. To facilitate comparison,  all $\Phi $'s are aligned such that the density maxima are positioned at site $x=15$. }
	\label{fig_fewbody_vbpvap}
\end{figure*}

The attractively interacting Bose-Hubbard model also provides a suitable setting for comparing and examining the performance of the PAV and VAP approaches. 

In Fig.~\ref{fig_fewbody_vbp}, we revisit the two cases in Fig.~\ref{fig_fewbody}, but with the PAV method. This time, we first get a single-configurational permanent $\Phi $ minimizing the energy [see Eqs.~(\ref{mconf})-(\ref{ratio})], and then project it onto each quasi-momentum subspace [see Eqs.~(\ref{projope})-(\ref{E3})]. The energy of the projected state and its overlap with the exact state are then calculated and displayed in Fig.~\ref{fig_fewbody_vbp}.  In the left panels, for $U=-1$, we see that the PAV method works very well, although not as well as the VAP method in Fig.~\ref{fig_fewbody}. The overlap deficit and relative energy error are both on the order of $10^{-3}$. In contrast, in the right panels, for $U=-0.5$, we see that the PAV method performs significantly worse. While the PAV results match the exact ones reasonably well for small values of $q$,  the discrepancy becomes  pronounced for larger $q$. There, the overlap deficit and relative energy error are both on the order of unity, indicating a breakdown of the PAV method in this regime.

It is expected that PAV works better for stronger attractions. In both approaches of PAV and VAP, the picture is that the pre-projection state $\Phi $ captures the internal structure, and the symmetry actions generate the collective motion. The key difference is that, in VAP,  $\Phi $ is allowed to adapt to the external motion, while in PAV, a common $\Phi $ is used across all symmetry sectors. For this to be a good approximation, the internal structure should be sufficiently rigid.  This condition is more likely to be fulfilled in the strong attraction limit. In this limit, the energy minimizing $\Phi $ corresponds to a tightly bound cluster localized on some site. Its excitation gap is large, and the coupling to other symmetry-related degenerate states is small due to its limited spatial extent. Both factors contribute to the rigidity of the cluster. 

To confirm this picture, in Fig.~\ref{fig_fewbody_vbpvap}, we study the $\Phi $'s in the VAP and PAV calculations in Fig.~\ref{fig_fewbody} and Fig.~\ref{fig_fewbody_vbp} by examining their density distributions $\rho(x,x) = \langle\hat{a}_x^\dagger \hat{a}_x \rangle $. For each value of $U$, we analyze the $\Phi $ from the PAV calculation and three representative $\Phi$'s from the VAP calculation. 
We see that for $U=-1$, the four distributions are all sharply peaked and closely resemble one another, while for $U=-0.5$, the four distributions are quite different, and in particular, the one associated with PAV is notably broader than before. This demonstrates that the internal structure is rigid for $U=-1$, but more flexible for $U=-0.5$, and accounts for the success  of PAV in the former case and its limitations in the latter. 

The lesson is that VAP outperforms PAV due to its use of a more adaptive internal wave function. Below, we shall not bother to present the PAV results any more as its performance is even poorer than in the current problem. 

\subsubsection{Repulsive case: $U > 0 $}

\begin{figure*}[tb]
	\centering
	\includegraphics[width= 0.48\textwidth ]{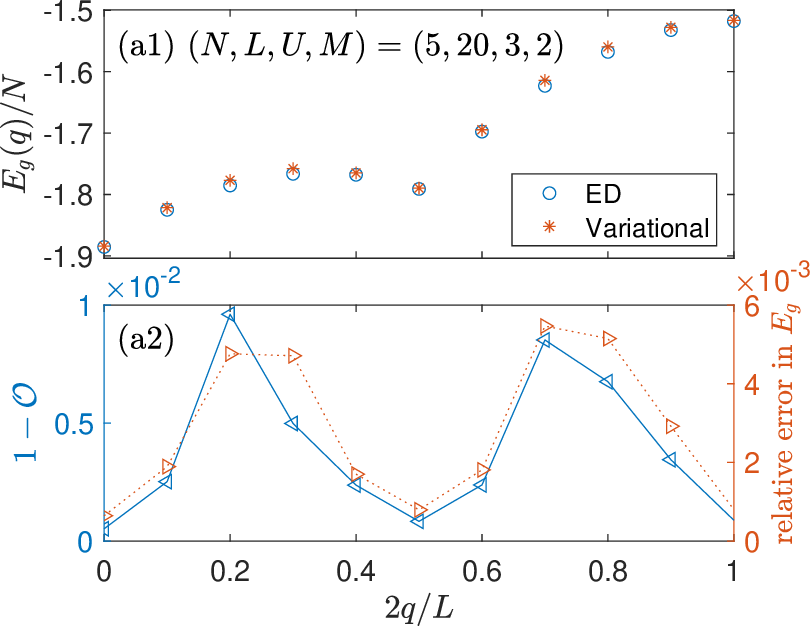}
	\includegraphics[width= 0.48\textwidth ]{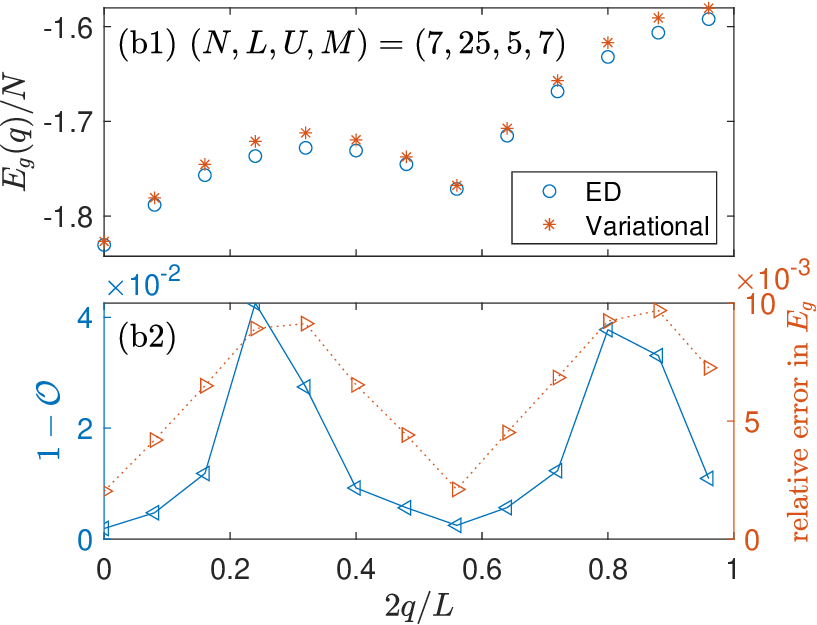}
	\caption{(Color online) Upper panels: Yrast line of a repulsively interacting ($U> 0$) Bose-Hubbard model (\ref{bhm}). For each quasi-momentum $0\leq q\leq \floor{L/2}$, we calculate the energy of the lowest eigenstate in the $q$-sector with both exact diagonalization (ED) and the projected variational wave function with $M$ configurations.  Lower panels: Overlap $\mathcal{O}$ between the exact state and the projected variational wave function (left y-axis), and the relative error in energy (right y-axis). }
	\label{fig_yrast}
\end{figure*}

In the repulsive case, we have a different picture than the attractive case, as the particles tend to disperse on the lattice instead of clinging to each other. We expect the Yrast line to be completely different from that in the attractive case. 

In Fig.~\ref{fig_yrast}, we study the Yrast line of some repulsively interacting Bose-Hubbard models both by exact diagonalization and by the variational method. Before discussing the results, some remarks concerning the particular cases of $q = 0$ or $L/2$ are in order. For $q=0$, as proven in the proceeding subsection, the parity of the lowest state is definite---It must be even. For $q=L/2$, however, the parity of the lowest state is uncertain as the proof above does not apply because now $U> 0 $. However, numerically, the observation is that the two lowest eigenstates are often nearly degenerate and with opposite parities. For the sake of definiteness, we choose to target the odd-parity one.

In both Fig.~\ref{fig_yrast}(a1) and Fig.~\ref{fig_yrast}(b1), we see that the Yrast line calculated with the projected variational wave function follows that with the exact diagonalization closely. 
A salient feature of the exact Yrast line is the dips or cusps, which occur when $q$ is an integral multiplier of $N $. The cusps were predicted for the continuum model partly because of the Galilean invariance
\cite{bloch}, but are also expected in a lattice model \cite{yang}. Here we see that they are nicely reproduced by the variational method. Actually, the variational data points are even closest to the exact ones at the cusps. Quantitatively, in Figs.~\ref{fig_yrast}(a2) and \ref{fig_yrast}(b2), we see that the relative error in energy and the deficit in overlap are on the order of $10^{-2}$ in the worst cases. Given the large Hilbert space, the interaction strength $U$, and the relatively small number of configurations $M$, the variational wave function performs remarkably well.

\subsubsection{Polaron problem}

\begin{figure*}[tb]
	\centering
	\includegraphics[width= 0.48\textwidth ]{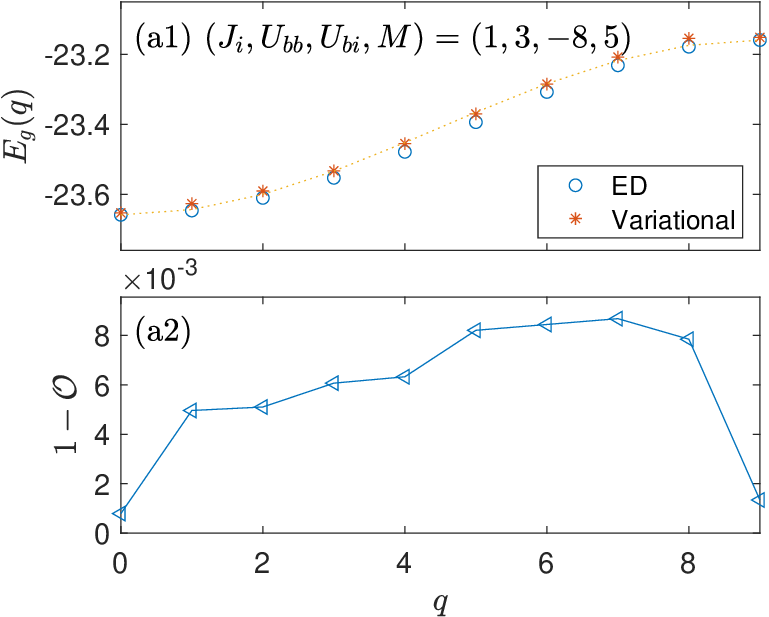}
	\includegraphics[width= 0.48\textwidth ]{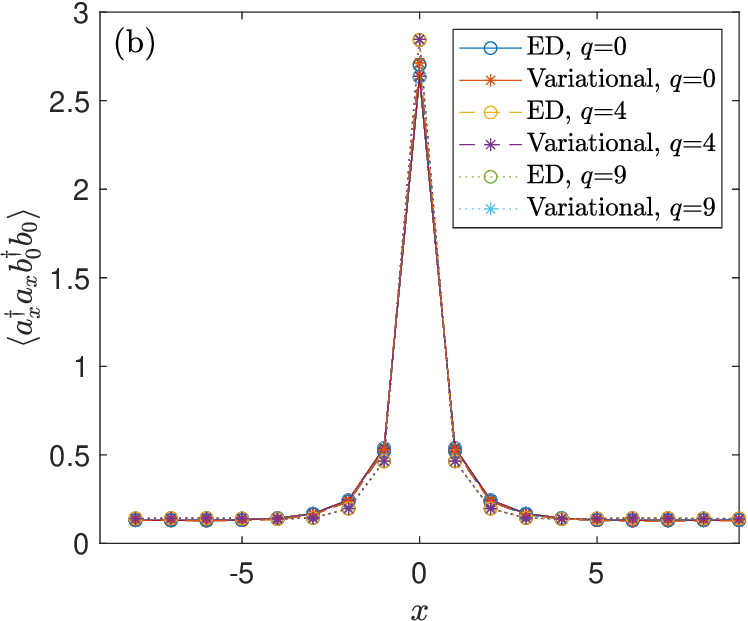}
	\caption{(Color online) Yrast state of the Bose polaron problem (\ref{polaronH}) with $N=6$ bath atoms and $L=18$ lattice sites. (a1) Energy of the lowest eigenstate with quasi-momentum $q$ calculated with both exact diagonalization (ED) and the projected variational wave function. (a2) Overlap between the exact state and the variational state. (b) Correlation function between the impurity atom and the bath atoms. Three states with different quasi-momenta are shown.  }
	\label{fig_polaron1}
\end{figure*}

\begin{figure*}[tb]
	\centering
	\includegraphics[width= 0.48\textwidth ]{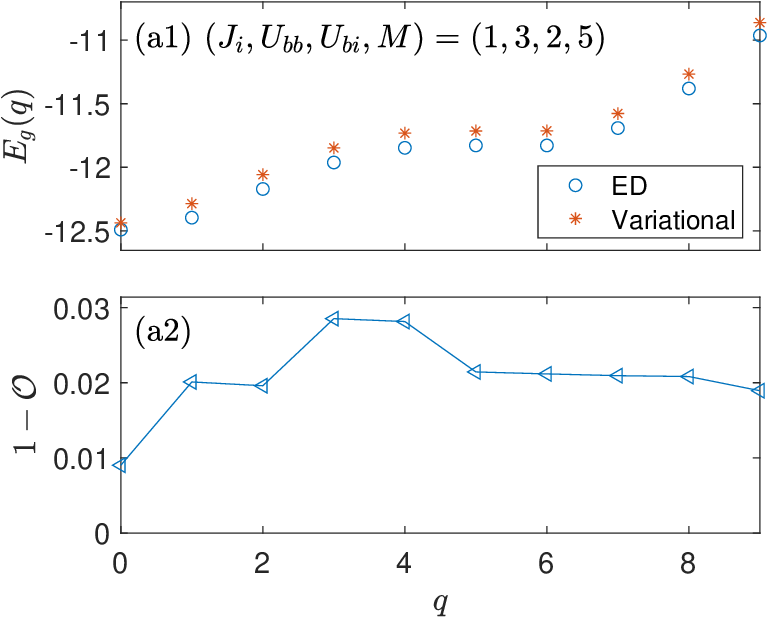}
	\includegraphics[width= 0.48\textwidth ]{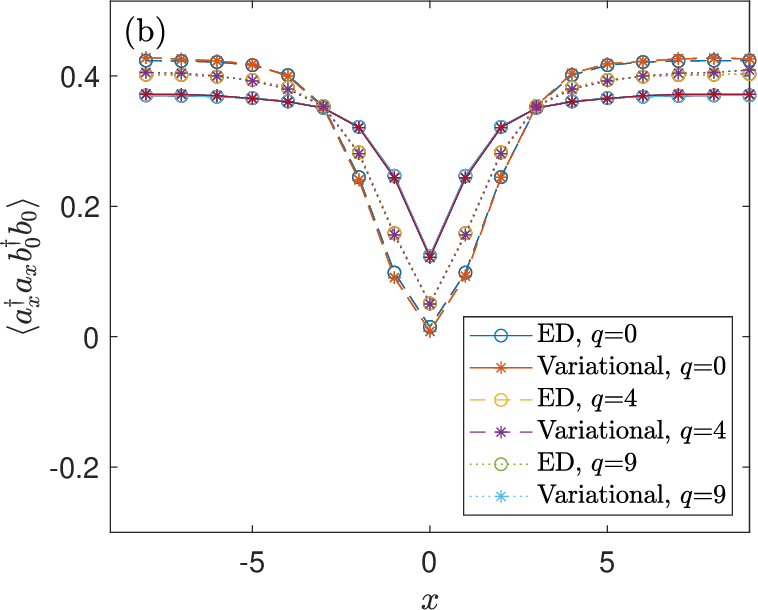}
	\caption{(Color online) Yrast state of the Bose polaron problem (\ref{polaronH}). (a1) Energy of the lowest eigenstate with quasi-momentum $q$ calculated with both exact diagonalization (ED) and the projected variational wave function. (a2) Overlap between the exact state and the variational state. (b) Correlation function between the impurity atom and the bath atoms. Three states with different quasi-momenta are shown.  }
	\label{fig_polaron2}
\end{figure*}

Another model realizing the $\text{D}_L$ symmetry is the Bose polaron problem. The scenario is that we add an impurity to the original Bose-Hubbard model. The Hamiltonian is 
\begin{eqnarray}\label{polaronH}
	\hat{H} &=& - \sum_{x=1}^{L} ( \hat{a}_x^\dagger \hat{a}_{x+1} +  \hat{a}_{x+1}^\dagger \hat{a}_{x})   + \frac{U_{bb} }{2} \sum_{x=1}^{L}  \hat{a}_x^\dagger \hat{a}_x^\dagger \hat{a}_x \hat{a}_x   \nonumber  \\
	&& - J_i\sum_{x=1}^{L} ( \hat{b}_x^\dagger \hat{b}_{x+1} +  \hat{b}_{x+1}^\dagger \hat{b}_{x})   + U_{bi}  \sum_{x=1}^{L}  \hat{a}_x^\dagger \hat{a}_x \hat{b}_x^\dagger \hat{b}_x .
\end{eqnarray}
Here the $a$-operators are for the bath atoms, while the $b$-operators are for the impurity atom. The parameter $J_i > 0 $ is the hopping strength of the impurity atom, $U_{bb}$ the on-site interaction between the bath atoms, while $U_{bi}$ the on-site interaction between the bath atoms and the impurity atom. As we have only one impurity atom, we have no impurity-impurity interaction term here. Depending on the sign of $U_{bi}$, the polaron is referred to as either an attractive polaron ($U_{bi} < 0$) or a repulsive polaron ($U_{bi} >  0 $). 

It is apparent that $\text{D}_L$ in (\ref{DL}) is the symmetry group. The only difference is that now $\hat{T}$ and $\hat{S}$ act on the bath atoms and the impurity atom simultaneously. 
We first classify the eigenstates by the quasi-momentum $q$. A total wave function of quasi-momentum $q$ is definitely of the form 
\begin{eqnarray}
	|\Psi^{(q)} \rangle = \sum_{n=0}^{L-1 } e^{-i 2 \pi q n /L } |n\rangle_b \otimes \hat{T}^{n} |\varphi \rangle_a . 
\end{eqnarray}
Here $|n\rangle_b = \hat{b}_n^\dagger |\text{vac}\rangle $ denotes the Fock state in which the impurity atom occupies the $n$th site, and $|\varphi \rangle_a$ denotes a state of the bath atoms.  Substituting this wave function into the eigenvalue equation, we get an equation for $|\varphi\rangle_a $, 
\begin{eqnarray}
	E |\varphi\rangle_a =\left [ \hat{H}_a  + U_{bi} \hat{a}_0^\dagger \hat{a}_0 -J_i (e^{-i 2 \pi q  /L } \hat{T} + e^{i 2 \pi q  /L }\hat{T}^\dagger) \right ] |\varphi\rangle_a.
\end{eqnarray}
If $q=0 $ or $L/2$, the effective Hamiltonian in the square brackets is invariant under the reflection $\hat{S}$. The eigenstates can then be divided into even-parity ($s =1$) ones and odd-parity ($s = -1$) ones. For $q=0$, again by the Frobenius-Perron theorem, it is easy to prove that the lowest state must be of even parity. For $q=L/2 $, the parity of the lowest state is undetermined. However, our experience is that if $U_{bi} < 0 $, the lowest state is often of even parity; while if $U_{bi} > 0 $, the lowest state is often of odd parity and there is an even-parity state nearly degenerate with it. We shall follow this empirical rule to target the lowest state in the $q=0$ or $q=L/2$ sectors. 

In Fig.~\ref{fig_polaron1}, we study an attractive polaron with $N = 6$ bath atoms on a lattice of $L =18 $ sites. The picture is that the bath atoms are attracted towards the impurity atom, and this tendency is countered by the repulsion between them. In the end, in a uniform background, a cloud of bath atoms dress the impurity atom and follow it as it itinerate on the lattice. The impurity plus the cloud as a whole is then called a polaron.  In Fig.~\ref{fig_polaron1}(a1), we see that the Yrast line is very close to a narrow-width cosine function. This indicates that, for the specific set of parameters, the polaron is a well-developed quasi-particle, in particular, a rigid and heavy one. This fact is also reflected in Fig.~\ref{fig_polaron1}(b), where we plot the correlation function between the impurity atom and the bath atoms, or more precisely, the density distribution of the bath atoms when the impurity atom is found at site $0$. We see that the distribution is almost independent of the quasi-momentum $q$. As for the performance of the variational wave function, in the three panels of Fig.~\ref{fig_polaron1}, we see that it has high overlap with the exact state ($0.99$ in the worst case), and delivers the energy and the correlation function accurately.

In Fig.~\ref{fig_polaron2}, we study the repulsive case. The only difference with Fig.~\ref{fig_polaron1} is that now the impurity-bath interaction is positive and  precisely,  $U_{bi } = 2 $. From Fig.~\ref{fig_polaron2}(a1) we see that, unlike in Fig.~\ref{fig_polaron1}(a1), now the Yrast line deviates significantly from a cosine function, which implies that the polaron cannot be pictured as a structureless particle hopping on the lattice.  Instead,  its internal and external motions are strongly coupled. This point is better revealed in Fig.~\ref{fig_polaron2}(b), where we see that the density distribution of the bath atoms around the impurity atom strongly depends on the quasi-momentum $q$. As for the performance of the variational wave function, as in Fig.~\ref{fig_polaron1}, it is doing very well. Its overlap with the exact state is 0.97 in the worst case and its correlation function nearly matches the exact result.  

\subsection{$\text{C}_L$ symmetry: Superfluidity}

\begin{figure*}[tb]
	\centering
	\includegraphics[width= 0.48\textwidth ]{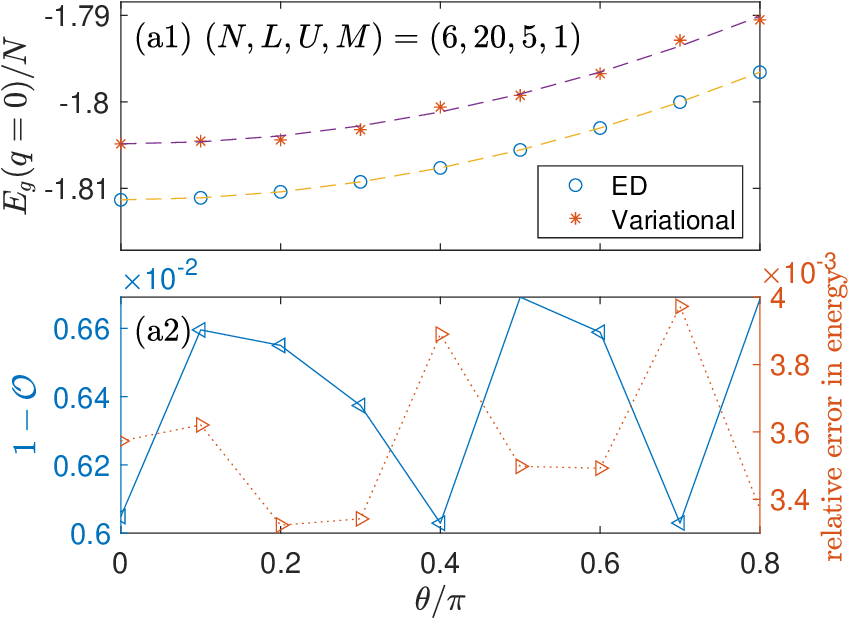}
	\includegraphics[width= 0.48\textwidth ]{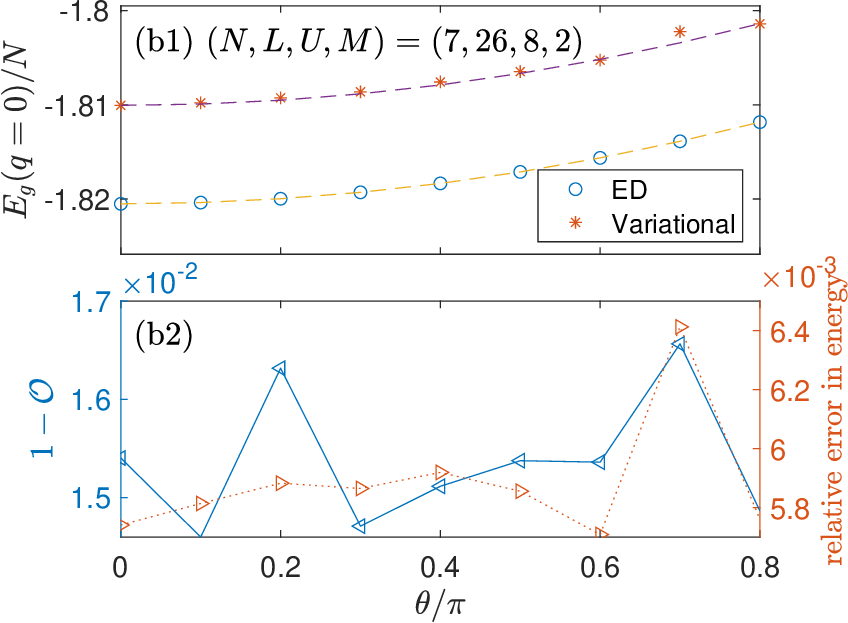}
	\caption{(Color online) Upper panels: Energy of the lowest eigenstate  in the $(q=0)$-sector as a function of the magnetic flux $\theta $, calculated with both exact diagonalization (ED) and the projected variational wave function with $M$ configurations. The dashed lines are guide of eye. They are graphs of some quadratic functions,  differing just by some shift in the vertical direction.  Lower panels: Overlap $\mathcal{O}$ between the exact state and the projected variational wave function (left y-axis), and the relative error in energy (right y-axis). }
	\label{fig_sf}
\end{figure*}

The $\text{C}_L$ symmetry is realized by a Bose-Hubbard ring threaded by a magnetic flux $\theta $. The Hamiltonian is 
 \begin{eqnarray}\label{bhm2}
	\hat{H} = - \sum_{x=1}^{L} (e^{i\theta /L} \hat{a}_x^\dagger \hat{a}_{x+1} + e^{-i\theta/L } \hat{a}_{x+1}^\dagger \hat{a}_{x})   + \frac{U }{2} \sum_{x=1}^{L}  \hat{a}_x^\dagger \hat{a}_x^\dagger \hat{a}_x \hat{a}_x .\quad
\end{eqnarray}
For a generic value of $\theta$,  the reflection symmetry of the original Bose-Hubbard model is broken, but the translation symmetry is preserved. The eigenstates are classified by the quasi-momentum $q$.  In each $q$-sector, the lowest eigenstate can be targeted by a projected variational wave function of the form 
\begin{eqnarray}\label{projectedwf2}
	\Phi^{(q)} = \frac{1}{L} \sum_{n=0}^{L-1} \sum_{\alpha=1}^M   e^{-i 2\pi q n /L }  \hat{\mathcal{S}}( \hat{T}^n \phi^{(\alpha)}_1, \ldots, \hat{T}^n \phi^{(\alpha)}_N ). 
\end{eqnarray}

We are particularly interested in the sensitivity of the ground state energy $E_g$ to the magnetic flux, as its second derivative with respect to $\theta $, $\partial^2 E_g /\partial \theta^2 |_{\theta = 0 }$, defines essentially the so-called superfluid fraction of the system \cite{gammal}. Now we note that when $\theta = 0 $, by the discussion above, the ground state is with $q = 0 $.  As $\theta $ is ramped up, the translation symmetry is preserved, so a state initially in a  $q$-sector will remain in this sector. By continuity, in a neighborhood of $\theta = 0 $, the ground state will remain in the $(q=0)$-sector. Therefore, we should set $q=0$ in the variational wave function above. 

In Fig.~\ref{fig_sf}, we compare the ground state energies and wave functions obtained by exact diagonalization and the variational method. We see that even with a single or just a few independent configurations, the projected variational wave function can get the energy to within a relative error on the order of $1\%$, and the deficit of overlap with the exact state is on the order of $1\%$. Furthermore, even though there is some error in energy, the response of the energy to the magnetic flux is the same, as demonstrated by the fact that the ED data and the variational data differ just by a global shift.  

\subsection{$D_4$ symmetry: a two-dimensional square lattice}

\begin{table}[tbh]
	\centering
	\begin{tabular}{c c c} 
		\hline
		$ $  & $\hat{T}$ & $\hat{S}$ \\
		\hline
		$D^{(1)} $ & $+1$ & $+1$ \\ 
		\hline
		$D^{(2)} $ & $+1$ & $-1 $\\ 
		\hline
	$	D^{(3)} $ & $-1$ & $+1$ \\ 
		\hline
	$	D^{(4)} $ & $-1 $ & $-1$ \\ 
		\hline
	$D^{(5)}$ & $\begin{pmatrix}
		0 & -1 \\
		1 & 0 
	\end{pmatrix} $ & $\begin{pmatrix}
	1 & 0 \\
	0 & -1
	\end{pmatrix} $ \\  
		\hline
	\end{tabular}
	\caption{The five irreducible representations of the diheral group $\text{D}_4 $.  Only the matrices corresponding to the two generators $\hat{T}$ and $\hat{S}$ are displayed.  }
	\label{tabled4}
\end{table}

\begin{figure*}[tb]
	\centering
	\includegraphics[width= 0.48\textwidth ]{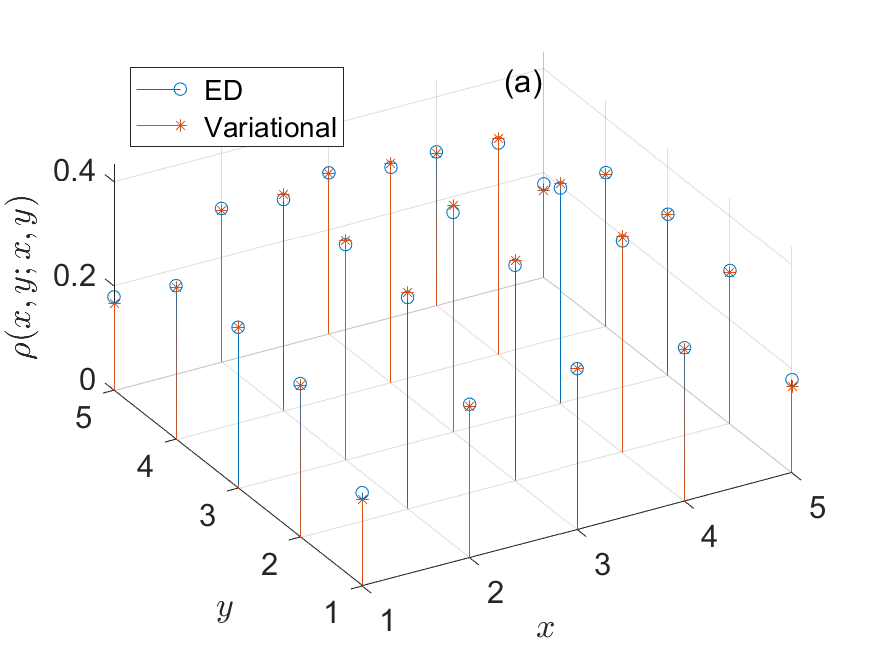}
	\includegraphics[width= 0.48\textwidth ]{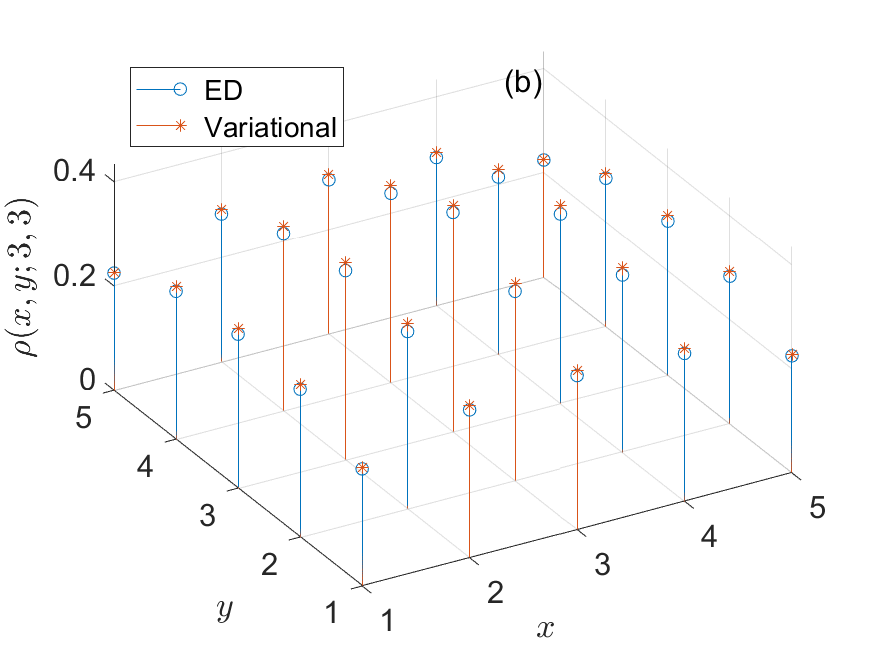}
	\caption{(Color online) (a) Density distribution $\rho(x,y;x,y)=\langle \hat{a}^\dagger_{x,y} \hat{a}_{x, y}\rangle $ and (b) off-diagonal correlator $\rho(x,y;3,3) = \langle \hat{a}^\dagger_{3,3} \hat{a}_{x,y } \rangle $ of the lowest eigenstate in the trivial representation of the $\text{D}_4$ group (see Table~\ref{tabled4}). The system is a Bose-Hubbard model on a two-dimensional square lattice with $L_x=L_y = 5 $. The particle number is $N =8$, and the on-site interaction strength is $U = 10$.  We have $M = 5 $ independent configurations in the projected variational wave function. We see that the variational results agree with those obtained by exact diagonalization (ED) very well, and enjoy the expected full symmetry of a square. The overlap between the exact state and the projected variational wave function is $\mathcal{O} = 0.9721$. The exact energy per particle is $E_{ext} =-2.7740 $, while the variational one is $E_{var} =-2.6950 $. }
	\label{fig_d4r1}
\end{figure*}

\begin{figure*}[tb]
	\centering
	\includegraphics[width= 0.48\textwidth ]{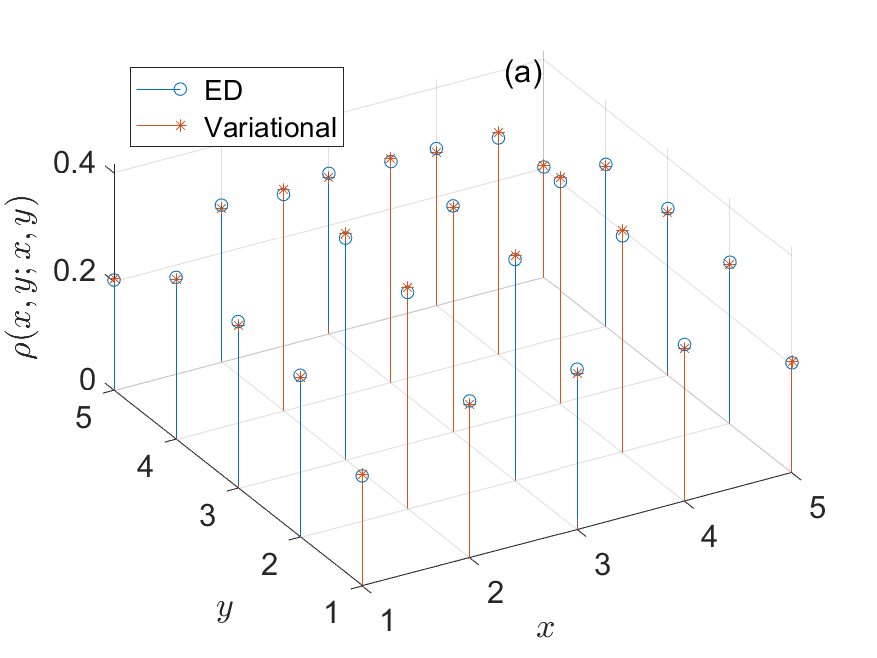}
	\includegraphics[width= 0.48\textwidth ]{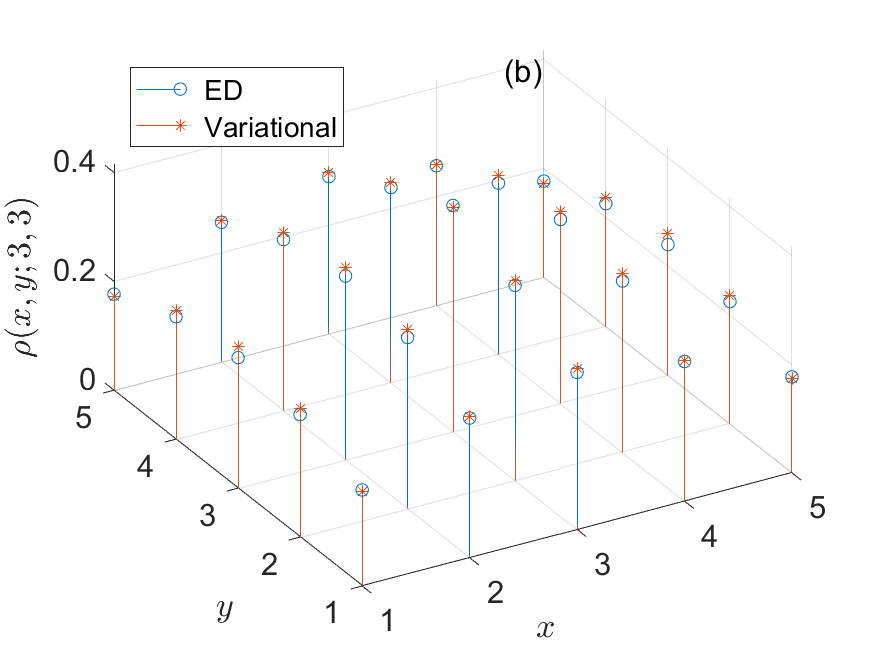}
	\caption{(Color online) (a) Density distribution $\rho(x,y;x,y)=\langle \hat{a}^\dagger_{x,y} \hat{a}_{x, y}\rangle $ and (b) off-diagonal correlator $\rho(x,y;3,3) = \langle \hat{a}^\dagger_{3,3} \hat{a}_{x,y } \rangle $ of the lowest eigenstate belonging to the first row of the 5th irreducible representation of the $\text{D}_4$ group (see Table~\ref{tabled4}). The system is the same as in Fig.~\ref{fig_d4r1}.  We have $M =5 $ independent configurations in the projected variational wave function. As in Fig.~\ref{fig_d4r1}, the variational results agree with those obtained by exact diagonalization (ED) very well. However, now the symmetry is lower. The distributions are only invariant under reflections along the $x-$ and $y$- axes. The overlap between the exact state and the projected variational wave function is $\mathcal{O} = 0.9478 $. The exact energy per particle is $E_{ext} = -2.6403$, while the variational one is $E_{var} = -2.5633$. }
	\label{fig_d4r5}
\end{figure*}

So far, we have considered only one-dimensional models. However, the variational method apparently applies in any dimension. Below we consider a two-dimensional square lattice with the open boundary condition. The Hamiltonian is 
\begin{eqnarray}
	\hat{H} =- \sum_{x= 1}^{L-1} (\hat{a}_{x,y}^\dagger \hat{a}_{x+1, y } + h.c.)- \sum_{y= 1}^{L-1} (\hat{a}_{x,y}^\dagger \hat{a}_{x, y+1 } + h.c.) + \frac{U }{2}\sum_{x,y=1}^L \hat{a}_{x,y}^\dagger  \hat{a}_{x,y}^\dagger \hat{a}_{x, y } \hat{a}_{x, y }. 
\end{eqnarray}
The symmetry group is the dihedral group $\text{D}_4 =\{\hat{T}^n , \hat{S}\hat{T}^n | 0\leq n \leq 3  \}$. Here $\hat{T}$ is the operation of rotation anti-clockwise by $90^\circ$ and $\hat{S}$ is the operation of reflection along the $x$-axis. They act on a single-particle wave function as $(\hat{T}\phi )(x,y) = \phi ( y, L+1 - x )$ and $(\hat{S} \phi )(x,y) = \phi(x, L+1 - y )$. In Sec.~\ref{dlsym}, we have actually constructed all the irreducible representations of a general dihedral group. For the specific group of $\text{D}_4 $, we have four one-dimensional irreducible representations and one two-dimensional representation. They are listed in Table.~\ref{tabled4}. 

In Figs.~\ref{fig_d4r1} and \ref{fig_d4r5}, we study such a square lattice Bose-Hubbard model with $(N,L,U )= (8,5,10 )$. In Fig.~\ref{fig_d4r1}, we target the ground state, which belongs to the representation $D^{(1)}$. In Fig.~\ref{fig_d4r5}, we target the lowest eigenstate belonging to the first row of the representation $D^{(5)}$. In both cases, we consider the density distribution $\rho(x,y;x,y)=\langle \hat{a}^\dagger_{x,y} \hat{a}_{x, y}\rangle $ and the off-diagonal correlator $\rho(x,y;3,3) = \langle \hat{a}^\dagger_{3,3} \hat{a}_{x,y } \rangle $, and for both we find good agreement between the variational state and the exact state. The overlap is 0.9721 and 0.9478, respectively. These numbers are impressive in view of the huge Hilbert space dimension ($\mathcal{D} = 10\;518\;300$) and very limited number of independent configurations ($M=5$).

\section{Conclusions and discussions}\label{secconclude}

We have developed a variation after projection scheme within the Hartree-Fock approximation for bosons. As a Hartree-Fock method, the building blocks of the variational wave function are Bose permanents. Starting from a set of $ M $ independent permanents, a symmetry-adapted wave function is constructed by the standard group theoretical method of projection.  It is again a combination of permanents, but generally with $|G|$-fold more permanents, and the permanents are related by the action of the elements of the symmetry group $G$. It is fortunate that the subsequent variation or optimization can be done in a manner essentially the same as in the pre-projection case. This method enables us to target the lowest eigenstate in any symmetry sector, or more precisely, the lowest eigenstate belonging to any row of any irreducible representation of the symmetry group. We have rigorously tested this approach on various models, across different dimensions, and with diverse symmetry groups.  In all instances, with a modest number of independent configurations, the symmetry-adapted variational wave function approximates the target wave function with high fidelity, as evidenced by the significant overlap between them. It then follows without surprise that the variational wave function can reproduce the energy and even the single-particle correlation functions accurately.  

The primary drawback of the method is that, because of the exponential time complexity of permanent computation, it can deal with only a handful of particles. That is, the method is only applicable to few-body systems. In this paper, the largest particle number is $N = 8 $ and in a previous paper \cite{zhang4}, this number is 12.  But from the point of view of few-body physics, these numbers are large enough. We note that at least on the few-body bound state problem, the $N=3 $ case is already challenging \cite{mattis, valiente} and the $N= 4 $ case is rarely studied in the literature. We also note that systems with 12 particles (or even less, like $N=6$) are sometimes referred to as mesoscopic, and can reveal precursors of many-body physics. In this regard, the best example is possibly the fractional quantum Hall systems, where a handful of particles ($N \lesssim 10$) can often capture the thermodynamic limit well (see Ref.~\cite{hu} and references therein).

Although the method is useful only for few-body or mesoscopic systems, it has several advantages whenever it is applicable:

(i) It can handle large-volume systems. The memory cost of the method is just storing the single-particle operators $\hat{F}$ and $\hat{G}$, which scales quadratically with the lattice size $L $ and is independent of the particle number $N $.  In contrast, in exact diagonalization (ED), one has at least to store a few vectors of the dimension of the Hilbert space, which scales with $L $ as $L^N $.  Therefore, as long as $N \geq 3 $, the current variational method can treat larger systems than ED. In this paper, as we have to compare the variational results with that of ED, we have not taken a too large $L $. But much larger $L $ can be treated. For example, with $N= 8 $ particles, we can at least take $L =100$, for which the dimension of the Hilbert space is as large as 325 billion, far beyond the capability of ED. 

(ii) The method allows us to target not only the global lowest eigenstate, which often belongs to the trivial representation of the symmetry group, but also local lowest eigenstates, i.e, lowest eigenstates in nontrivial representations of the symmetry group. In this regard, it is superior to many quantum Monte Carlo methods, which generally can only target the global lowest eigenstate. Moreover, unlike many quantum Monte Carlo methods, the method is free of  the sign problem and can deal with complex hoppings as in the study of superfluid fraction. 

(iii) The method also works in any dimension and with both short- and long-ranged interactions.  Although in this paper, we have considered at most two dimensions and only on-site interactions, the method generalizes straightforwardly to higher dimensions and long-ranged interactions (say, dipole-dipole interaction as in \cite{will}). Moreover, we see no reason that the method cannot work equally well in these cases. 

The fact that a bosonic wave function comprising millions of components can be well approximated by a very limited number of permanents demonstrates the advantage of working with non-orthogonal single-particle orbitals. By sacrificing orthogonality, we gain compactness and save memory costs, enabling us to treat larger systems. This approach should also be applicable to fermions. We can attempt to approximate a fermionic wave function with a finite number of non-orthogonal Slater determinants. Although the single-particle orbitals used to construct each determinant can always be orthogonalized, the orbitals of different determinants do not need to be orthogonal. In other words, we do not have a common orthonormal single-particle basis for the determinants. For a given target fermionic wave function and a fixed number of determinants, the overlap can be maximized using the greedy strategy  too. Preliminary numerical findings indicate that for three fermions in six orbitals (the so-called Borland-Dennis setting \cite{borland1,borland2}), a generic fermionic wave function can always be expressed as the sum of \emph{two} determinants \cite{unpublished}. This result is intriguing. On the one hand, this number is minimal as a generic wave function is not a Slater determinant, and we need at least two; on the other hand, that this number is sufficient is surprising, as the Hilbert space dimension is as large as twenty.

The permanent variational wave functions have high overlap with the target states. However, these variational states are obtained by minimizing the energy instead of maximizing the overlap. Therefore, the overlap could be even higher if the objective is to maximize it. Here arises then a problem concerning the structure or complexity of many-particle wave functions \cite{ojanen}. A generic bosonic (fermionic) wave function, which is not a permanent (determinant) state, can be approximated to what extent by a combination of permanents (determinants)? More precisely, what is the maximal value of the overlap? In the fermionic case, we have the well-established concept of optimal Slater approximation \cite{aoto,zhang1, zhang2, zhang3}, which has proven useful in addressing many problems, such as the study of many-body localization \cite{mbl}. Similarly, in the bosonic case, we can introduce the notion of optimal permanent approximation. Whether this concept will be useful awaits further investigation. 

From the mathematical perspective, the problem here involves tensor decomposition or tensor approximation. A bosonic wave function is a symmetric tensor and the problem is to approximate it with a combination of some simple symmetric tensors, namely permanents. It is noteworthy that in the mathematical literature \cite{landsberg, siam1, siam2}, a similar problem has been extensively studied, where the permanents are further constrained to the simplest form, i.e., with all the orbitals identical. In physical terms, the building blocks the mathematicians use are condensate-type wave functions that arise in the Gross-Pitaevskii approximation. They have the important notion of symmetric tensor rank, which is the minimal number of condensate wave functions needed to recover the original state. In our case, we would be more interested in the minimal number of general permanents for this purpose. Undoubtedly, general permanents are more effective in reproducing the target state. While it is beyond our comprehension why mathematicians have focused only on the restricted case, hopefully our current problem can provide some motivation for the study of the more general case. 

In this paper, we have focused on static properties, specifically the lowest eigenstates. The permanent variational wave function can approximate these static objects with high accuracy. A natural question then arises, if we allow the single-particle orbitals to be time-dependent, can this approach also effectively approximate or follow a time-evolving state?  It is important to note that even tracking the time evolution of a few-body system can be highly challenging. The conceptually simplest approach is full diagonalization of the hamiltonian, but this approach is severely constrained by the size of the Hilbert space. A more sophisticated technique involves approximating the exact state with a matrix product state and attempting to evolve it \cite{mps}. However, this method faces difficulties due to the growth of entanglement and is typically limited to one-dimensional systems. In contrast, the permanent wave function approach approximates the exact state in a fundamentally different manner, allowing it to avoid these issues. It should be worthwhile to explore this.


\section*{Acknowledgments}

The authors are grateful to J. Guo, K. Yang, Y. Xiang and K. Jin for their helpful comments.
This work is supported by the National Key R\&D Program of China (No.
2021YFB3501503),  the Special Fund for
Research on National Major Research Instrument (No.
62327804) and the Foundation of LCP.


%




\begin{thebibliography}{99}
\bibitem{comment1}
Actually, as the orbitals in a permanent are generally non-orthogonal, they cannot be simultaneously the eigenstates of a hermitian operator.
\bibitem{n2case}
In the two-particle ($N=2$) case, $\hat{G}$ is just positive semi-definite and can have zero eigenvalues. But this does not cause much difficulty and is easy to handle. See Sec.~4.2 of Ref.~\cite{zhang4} for details.
\bibitem{numerical}
For us, the $\hat{F}$ and $\hat{G}$ operators are purely numerical entities (matrices actually). They generally do not have a simple analytic expression.
\bibitem{linear}
It is worth noting that here we have a linear equation for the orbital, while in the Gross-Pitaevskii approximation the equation for the optimal orbital is a nonlinear one.
\end{thebibliography}


\end{document}